\documentclass[aps,prl,amsmath,amssymb,showpacs,showkeys,reprint]{revtex4-1}
\usepackage[utf8]{inputenc}
\usepackage{amsmath}
\usepackage{graphicx}
\usepackage{dcolumn}
\usepackage{bm}
\begin{document}
\title{Matterwaves, Matterons, and the Atomtronic Transistor Oscillator}
\author{Dana Z. Anderson}
\affiliation {ColdQuanta Inc. 3030  Sterling Circle, Boulder CO 80301, USA}
\affiliation {Department of Physics and JILA, University of Colorado, Boulder, Colorado, 80309-0440, USA}

\preprint{APS/123-QED}

\begin{abstract}
A self-consistent theoretical treatment of a triple-well atomtronic transistor circuit reveals the mechanism of gain, conditions of oscillation, and properties of the subsequent coherent matterwaves emitted by the circuit.  A Bose-condensed reservoir of atoms in a large source well provides a chemical potential that drives circuit dynamics.  The theory is based on the ansatz that a condensate arises in the transistor gate well as a displaced ground state, that is, one that undergoes dipole oscillation in the well. That gate atoms remain condensed and oscillating is shown to be a consequence of the cooling induced by the emission of a matterwave into the vacuum.  Key circuit parameters such as the transistor transconductance and output current are derived by transitioning to a classical equivalent circuit model.  Voltage-like and current-like matterwave circuit wave fields are introduced in analogy with microwave circuits, as well as an impedance relationship between the two.  This leads to a new notion of a classically coherent matterwave that is the dual of a coherent electromagnetic wave and which is distinct from a deBroglie matterwave associated with cold atoms. Subjecting the emitted atom flux to an atomic potential that will reduce the deBroglie wavelength, for example, will increase the classical matterwave wavelength.  Quantization of the classical matterwave fields leads to the dual of the photon that is identified not as an atom but as something else, which is here dubbed a ``matteron".  
 \end{abstract}
 
 \date{\today}
\maketitle

\section{Introduction}
\begin{figure}[t]
\includegraphics[width=\columnwidth]{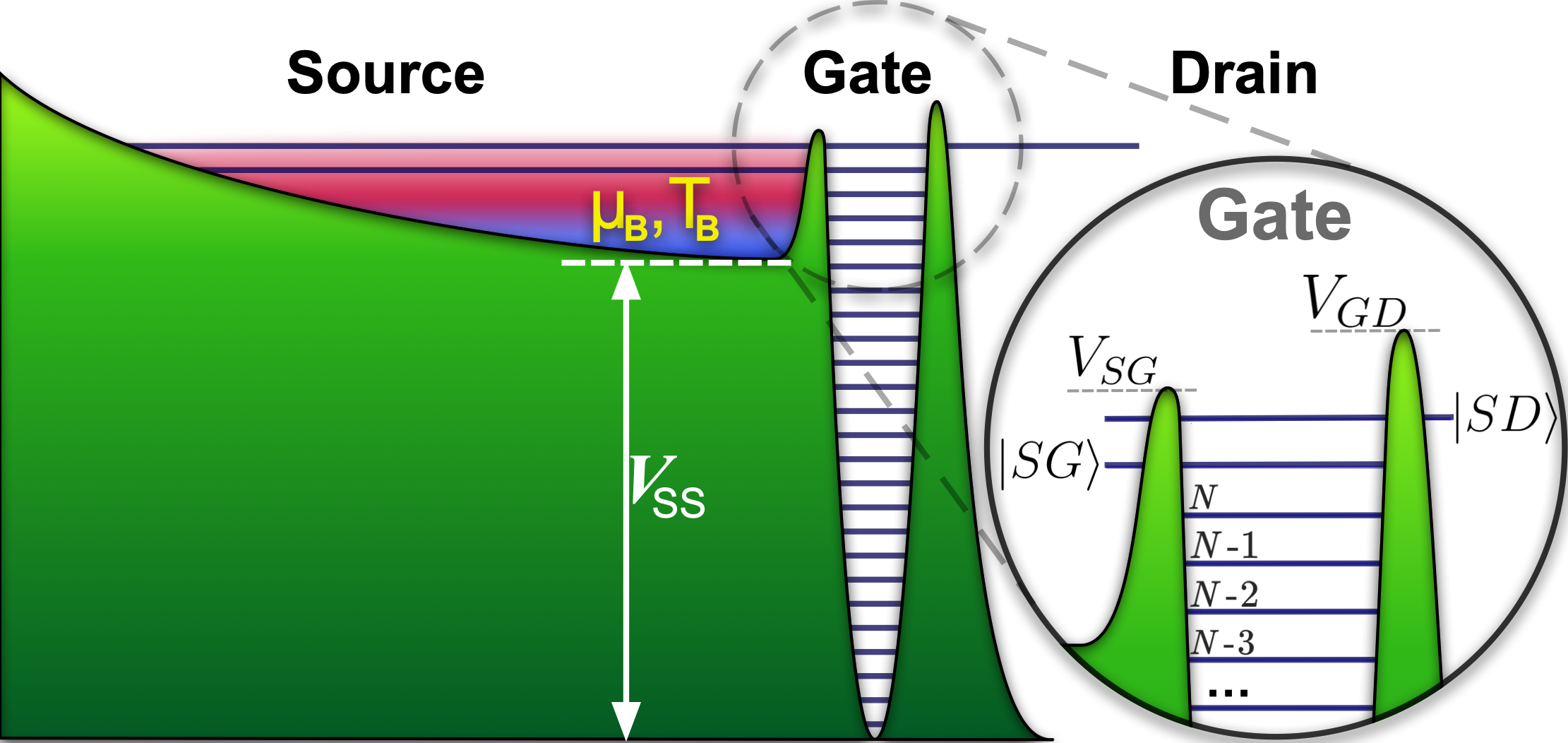}
\caption{\label{fig:BiasedTransistor} The matterwave oscillator circuit consists of a triple-well atomtronic transistor having a large bias $V_{SS}$ applied to its source well.  The circuit is powered by a ``battery" consisting of an ensemble of particles residing in the source well at a specified chemical potential $\mu_{\rm{B}}$ and temperature $T_{\rm{B}}$. The oscillation frequency is determined by the gate harmonic oscillation frequency; in addition to this frequency, the gate is characterized by the number of fully trapped levels $N$. Circuit feedback is set by the barrier height difference. The drain well is at vacuum. In this figure the vertical direction depicts potential energy and the horizontal direction is spatial coordinate.    }
\end{figure}

\begin{figure}
\includegraphics[width=\columnwidth]{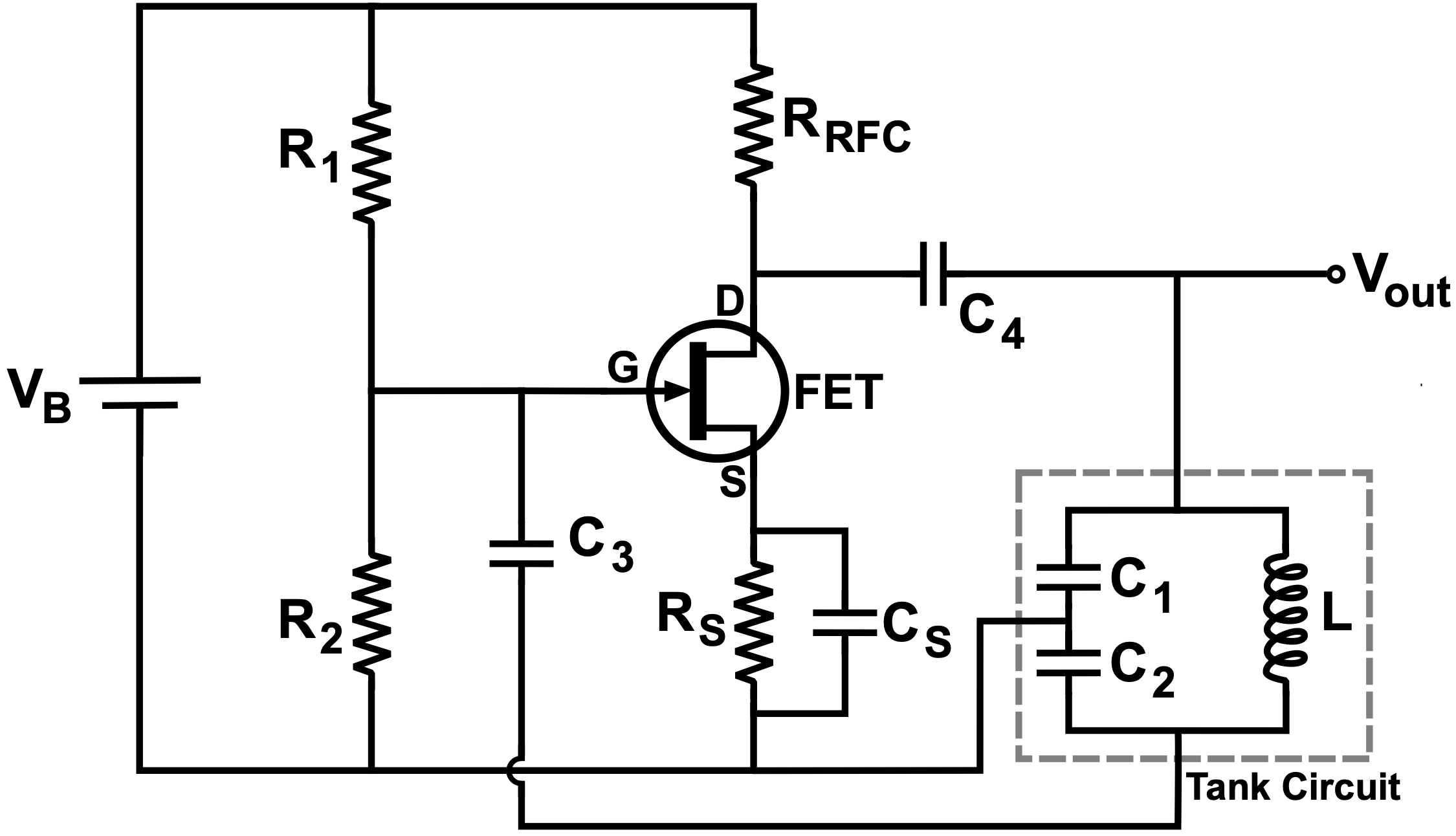}
\caption{\label{fig:ColpittsFET} The textbook single-transistor electronic Colpitts oscillator has much in common with the matterwave oscillator, keeping in mind that electric current is opposite to the direction of the flow of electrons, although elements in the matterwave version are distributed rather than bulk as suggested by a schematic.  In particular the LC "tank" circuit is schematic of the resonant frequency of the atomtronic transistor gate well while the gate bias voltage established by the resistor voltage divider $R_1$ and $R_2$ is schematic of the bias applied to the source well for the atomtronic circuit.  The dual to the feedback provided from the tank circuit to the FET gate corresponds to the difference in barrier heights of the atomtronic transistor.} 
\end{figure}

This work seeks to elucidate the principles of an atom-based transistor oscillator ``circuit" \cite{caliga2012}.  The circuit, which is illustrated in Fig. \ref{fig:BiasedTransistor}, incorporates a triple-well  transistor potential; namely an atomic potential consisting of a broad ``source" well, a narrow ``gate" well and a flat ``drain" well \cite{stickney2007,caliga2016b,caliga2016a}.  The source well sits at a bias potential $V_{\rm{SS}}$ relative to the other two wells, and contains a Bose-condensate. As a many-body system operating in non-thermal equilibrium the microscopic circuit physics is complicated, yet the qualitative behavior can be captured by analogy with its electronic counterpart \cite{seaman2007,pepino2009}.  The objective of this work is to extract an equivalent circuit out of the physics, from which the circuit's behavior in a classical regime can be determined and pedagogically understood.  Indeed, the equivalent circuit is schematically nearly identical to one that embodies an electronic transistor oscillator, such as the Colpitts oscillator shown in Fig. \ref{fig:ColpittsFET} \cite{azadmehr2020}.  

Despite the classical viewpoint, however, we will find that the atom current into the drain should \textit{not} be pictured as a stream of atoms, say, with a flux that oscillates in time as one might picture the electrons in an electronic oscillator.  Rather, the atomtronic oscillator emits a coherent matterwave into the drain in much the same manner that an electronic oscillator emits a coherent electromagnetic wave when the circuit is appropriately coupled to the vacuum utilizing an antenna.  The emitted matterwave is entirely distinct from the customary deBroglie wave that characterizes massive particles.

In driving toward an equivalent circuit it is our intent to bring to light the quantum origin of transistor gain and to capture the essential aspects needed to understand the driven oscillator dynamics. The equivalent circuit utilizes atomtronic duals of conventional electronic circuit parameters, such as the transconductance of the transistor. The use of circuit concepts and utilizing electronic duals allows one to leverage the principles and heuristics that are well known in electronic circuits in order to understand an otherwise complicated quantum system.

The evolving literature of atomtronics thus far addresses three distinct aspects of the atomic analog of electronics.  One involves  atom analogs to superconducting devices and circuits, such as Josephson junctions and SQUIDS \cite{ryu2007,ramanathan2011,wright2013,gallemi2015,aghamalyan2015,mateo2015,gallemi2016,tengstrand2021}(though it is perhaps notable that atomtronic circuitry does not require cryogenics to operate).  These atomtronic circuits operate as closed, i.e., energy conserving, systems and are interesting from a fundamental viewpoint as well as for their potential in sensing applications.  A second involves atomtronic circuits and devices operating as a lumped element system, meaning one in which the relevant wavelengths are typically long compared with the size of the various circuit elements, and/or are used to investigate mass transport, heat transport, and related phenomena \cite{benseny2010,filippone2014,li2016,polo2016,krinner2017,you2019,benseny2019,impens2020}. 

As indicated above, our case deals with an open quantum system in non-thermal equilibrium and our interest is in the matterwave properties of the circuit.  Energies of interest correspond to free-space matterwave wavelengths on the order of $1\mu\rm{m}$ and less. Thus as an analog of electronics our context is most closely that of microwave circuitry in which wavelengths are smaller than or on the order of the length scales associated with the components. Microwave circuit design makes extensive use of wave impedance concepts (highlighted by the prevalent use of Smith charts).  Impedance concepts are applicable and extremely informative in understanding atomtronic transistor circuits as well.  In this, what we can refer to as the microwave regime, atomtronic circuits are of interest for their practical potential in quantum signal processing, quantum sensing, and other information processing contexts. 

\section{Overview}
The thrust of our theoretical development begins with a  treatment of the gate well as a closed system having a harmonic particle potential.  Such a closed system in thermal equilibrium at sufficiently low temperature consists of a Bose-condensed gas in the ground state. With the coupling of the gate to the source and drain wells through barrier tunneling, one is driven to consider the dynamics of a system in non-thermal equilibrium.

This work's pivotal ansatz is that the dynamics of the system is such that a condensate forms not in the ground state of the gate well, but in a displaced ground state, i.e., a coherent state of the gate harmonic oscillator. This ansatz cannot be justified in isolation, but rather is justified through a self-consistent solution to circuit dynamics. 

The transition from a closed to an open quantum system treats the gate as coupled via tunneling to a reservoir of particles at fixed temperature and chemical potential comprising a ``battery" associated with the source well on the one side \cite{zozulya2013,caliga2017}, and to the vacuum on the other side. The battery drives the circuit dynamics.  A many-body approach is used to analyze the interaction energy between the harmonic oscillator modes of the gate with those that couple the gate to the two other wells. Subsequently, though, we move to the classical limit in order to most easily capture the open-systems character of both the battery and the vacuum.  We utilize the insight offered by an equivalent circuit model to impose self-consistency of a stationary solution for the circuit dynamics.  Transistor gain is seen to arise from a phase-dependent dynamical particle potential.  We establish that coupling to the oscillating condensate (or dipole) gives rise to the emission of a coherent matterwave into the drain well.  Self-consistency provides circuit currents and potentials, as well as the characteristics of the emitted matterwave. 

Taking the presence of the gate coherent state as an ansatz, it should be said that we do not here establish how the coherent state arises in the first place, nor are we in a position to carry out a formal stability analysis of the solution. For example, in conjunction with the gate coherent state ansatz is an underlying assumption that the (many-body) state of the gate can be written as a direct product, allowing us to replace particle operators with c-numbers and facilitating the transition to a classical circuit model. In doing so we neglect aspects such as the self-interaction terms among gate particles, hypothesizing that though they may have quantitative impact, they do not impact the qualitative behavior and properties of the system.  Whether or not such interactions lead to significant breakdown of the direct product assumption is a question that goes beyond the scope of this work. In exchange for its simplifying assumptions, however, the self-consistent approach provides clear predictive outcomes that are eminently verifiable through experiments, such as the properties of emitted matterwaves, thermal effects in the circuit, and oscillation thresholds.

Before delving into the heart of the theory, the next section reviews key lessons from a semi-classical kinetic treatment of the transistor that hint at the quantum behavior. Following that, an analysis of the oscillator begins by defining a model for the atomtronic transistor circuit. 

\section{Conclusions from Kinetic Theory}
Three results from an earlier kinetic treatment foreshadow the quantum behavior as well \cite{{caliga2016b,caliga2016a}}. The model teaches us that the behavior of the device is critically governed by a feedback parameter $\upsilon=(V_{\rm{GD}}-V_{\rm{GS}})/(k_{\rm{B}}T_{\rm{B}})$, where $V_{\rm{GD}}$,  $V_{\rm{GS}}$ are the barrier heights, $k_{\rm{B}}$ is Boltzmann's constant, and $T_{\rm{B}}$ is the temperature of the particles in the source well. Current will flow from the source to the drain provided that the barriers are sufficiently low and/or sufficiently narrow. An unintuitive result from the kinetic treatment is that atoms in the gate are \emph{colder}, $T_{\rm{g}}<T_{\rm{B}}$, and acquire a higher chemical potential, $\mu_{\rm{g}}>\mu_{\rm{s}}$, than those in the source given feedback above a particular threshold value. It is not a coincidence that a reverse potential drop also plays an essential role in electronic transistor operation.  Indeed, for a range of values of the critical parameter the reverse potential increases with current, corresponding to a negative resistance, which is a signature of device gain.   We note that the spontaneous formation of a Bose-condensate in the gate underlies the theory's prediction of a reverse-bias for the atomtronic transistor. 

Other than the appearance of the Bose-condensate in the gate, the kinetic theory cannot provide deeper insight into the nature of gain, and in particular possible quantum aspects of the current flow.  This work presents a variation of our earlier triple-well device in which a large bias $V_{\rm{SS}}$ is added to the source well and the source atoms are made to be \textit{very} cold. There are two energy scales that determine our meaning of ``very cold": one is the range of energies $\delta E$ over which atom transport across a single barrier is non-classical, that is, wave-like. Given an ensemble of atoms with thermal energy $k_{\rm{B}} T\lesssim\delta E$ on one side of a barrier, the flux of atoms to the other side will be dominated by the non-classical component.   The other is the characteristic energy level spacing $\Delta E_{\rm{G}}$ of the gate.   When powered by atoms having very low temperature, i.e., such that $\Delta E_{\rm{G}}>k_B T$, we can expect the behavior of the atomtronic transistor to be dominated by quantum effects.

\section{Transistor Model}
Our treatment of the oscillator circuit considers the source well as providing access to a supply of bosonic particles (atoms) characterized by temperature $T_{\rm{B}}$ and chemical potential $\mu_{\rm{B}}$. The supply is presumed to be sufficiently large as to not be depleted so that $T_{\rm{B}}$ and $\mu_{\rm{B}}$  can be taken as fixed over time scales of interest.  The energy of these particles sits on top of a bias potential $V_{SS}$.  The drain well is simply the vacuum, extending infinitely on to the right in Fig. \ref{fig:BiasedTransistor}.  
Our treatment of the transistor considers the gate well as parabolic, having a set of single-particle modes 
$\left\{ \left| i \right\rangle ,i = 0,1, \cdots N \right\}$ confined to the gate region, i.e., tunneling to the source or drain is taken to be negligible. To simplify our treatment we consider only an additional two modes that have significant probability for being found in the source well, $\left|N+1 \right\rangle \equiv \left| \rm{SG}\right\rangle$ and $\left| N+2\right\rangle \equiv \left| \rm{SD}\right\rangle$. Furthermore, given that the gate-drain barrier is higher than the gate-source barrier, i.e., the feedback parameter $\upsilon$ is positive, we consider that only the upper state, $\left| \rm{SD}\right\rangle$, of the two has significant coupling to the drain well.  Let us refer to $\left| \rm{SG}\right\rangle$ and $\left| \rm{SD} \right\rangle$ as the \textit{transistor modes}.

We utilize a many-body treatment in these next four sections specifically to capture the interaction energy between the gate and transistor modes. Since the gate well is parabolic, its energy levels have a uniform spacing $\Delta E_{\rm{G}}=\hbar\omega_0$, where we refer to $\omega_0$ as the \textit{gate frequency}. From here-on we assume that that the source particles are very cold, $k_{\rm{B}}T_{B}/\hbar \omega_0< 1$. The drain is characterized by a continuum of modes, $\{\left|\omega\right>\}$.

Transistor dynamics is governed by a Hamiltonian which we separate into that describing the transistor, gate, and drain modes, and their interactions. 
\begin{equation}
	\hat H=\hat H_{\rm{T}}+\hat H_{\rm{G}}+\hat H_{\rm{D}}+\hat H_{\rm{DT}}+\hat H_{\rm{GI}}.
\end{equation}
Associated with the discrete gate modes are a set of particle creation and annihilation operators:
\begin{equation}
\hat b_{i}, \hat b_{i}^{\dagger} 	\{i=0,1,2, ...N\} 	,
\end{equation}
while we associate creation operators $\hat b_{\rm{SG}}^{\dagger}$ and $\hat b_{\rm{SD}}^{\dagger}$ with the transistor modes, along with corresponding annihilation operators. The creation and annihilation operators obey the standard commutation relations:
\begin{equation}
\begin{split}
\left[\hat b_i,\hat b_i\right]& =[\hat b_i^\dagger,\hat b_i^\dagger] =0 \\
[\hat b_i,\hat b_j^\dagger]&=\delta_{ij},
\end{split}
\end{equation}
where $\delta_{ij}$ is the Kronecker delta. In terms of the particle operators the gate and transistor Hamiltonians are:
\begin{equation}\label{Eq:SHOHamiltonian}
	\hat H_{\rm{G}} = \hbar \omega_0\sum_{j=0}^{N} \left(j+\frac{1}{2} \right) 
	\hat b_{j}^\dagger \hat b_{j},
\end{equation}
and
\begin{equation}
		\hat H_{\rm{T}}=\hbar\omega_0\left[ \left(N+\frac{3}{2}\right)\hat b_{\rm{SG}}^\dagger \hat b_{\rm{SG}}+\left(N+\frac{5}{2}\right)\hat b_{\rm{SD}}^\dagger \hat b_{\rm{SD}} \right].
\end{equation}
$\hat H_{\rm{DT}}$ characterizes tunneling of the upper transistor mode into the continuum of the drain. The final contribution to the Hamiltonian arises from the interactions among the particles within the gate: 
\begin{equation}\label{Eq:ManyBodyHamiltonian}
\hat H_{\rm{GI}}=\frac{\eta}{2}\sum_{ijkl=0}^{N+2} U_{ijkl} \hat b_{i}^{\dagger}\hat b_{j}^{\dagger}\hat b_{k}\hat b_{l} ,
\end{equation}
where the overlap coupling factor is:
\begin{equation}
	U_{ijkl} = \int {\psi _{i}^ *(x) \psi _{j}^ *(x)\psi _{k}(x)\psi _{l}(x) dx},
\end{equation}
$\eta$ is the coupling strength:
\begin{equation}
	\eta = \frac{4\pi}{A} \frac{{\hbar ^2}{a_s}}{m },
\end{equation}
$a_{s}$ is the scattering length, $m$ is the particle mass, and $A$ is an effective cross-sectional area associated with the otherwise one-dimensional modes of the system.

\section{Transistor States}
The semi-classical treatment of the transistor highlights the importance of the feedback parameter on the system dynamics —namely it determines the threshold for condensate formation in the gate.  In the quantum case the analogous parameter is the temperature-normalized level spacing $\hbar \omega_0/k_{\rm{B}}T_{\rm{B}}$.  In particular we shall see that coupling of the transistor modes induced by an oscillating condensate leads to gate cooling.  Indeed, the presence of tunneling is ultimately responsible for oscillation of the gate condensate.  In addition to tunneling, the transistor modes also provide the mechanism for coupling of atoms in the source with the gate.  For the sake of numerical estimates, it is convenient to suppose that in the region of the gate the wavefunctions are just those harmonic oscillator wavefunctions corresponding to an isolated harmonic potential.  To this end, and in absence of tunneling we can write:
\begin{eqnarray}
	\left|\rm{SG}\right > \rightarrow \sin(\theta_1)\left|S_1\right>+\cos(\theta_1)\left |N+1\right>\\
	\left|\rm{SD}\right > \rightarrow \sin(\theta_2)\left|S_2\right>+\cos(\theta_2)\left |N+2\right>,
\end{eqnarray}
in which $\left|N+1\right>$ and $\left|N+2\right>$ are the $(N+1)$ and  $(N+2)$ modes of the gate harmonic oscillator while $\left|S_1\right>$ and $\left|S_2\right>$ are orthogonal modes that primarily occupy the source well.  Since the source and gate are separated by a barrier through which particles must tunnel, in reality each of the above source modes are split into two - a symmetric and an anti-symmetric pair.  However, we suppose that the tunneling energy is small and can be neglected for our purposes. It is furthermore convenient for pedagogy, without sacrifice of the qualitative behavior, to suppose that the mixing angles are equal, $\theta_1=\theta_2\equiv\theta$.
\begin{eqnarray}
	\left|\rm{SG}\right > = \sin(\theta)\left|S_1\right>+\cos(\theta)\left |N+1\right>\\
	\left|\rm{SD}\right > = \sin(\theta)\left|S_2\right>+\cos(\theta)\left |N+2\right>.
\end{eqnarray}
The presence of an interaction suggests that it will be useful to work in a \textit{normal mode} basis for the transistor modes.  We introduce symmetric and anti-symmetric creation operators:
\begin{equation}\label{Eq:SymmetryOperators}
\hat b_ \pm ^\dag  = \frac{1}{\sqrt 2 }\left( {\hat b_{\rm{SD}}^\dag  \pm \hat b_{\rm{SG}}^\dag } \right),
\end{equation}
and similarly for the annihilation operators.  We designate eigenstates of these operators as:
\begin{eqnarray}
	\label{Eq:Symmetric} \hat b_{+}\left|+\right\rangle=\sqrt{M_+}\left|+\right\rangle\\
	\label{Eq:AntiSymmetric}  \hat b_{-}\left|-\right\rangle=-\sqrt{M_{-}}\left|-\right\rangle.
\end{eqnarray}

\begin{figure}\includegraphics[width=\columnwidth]{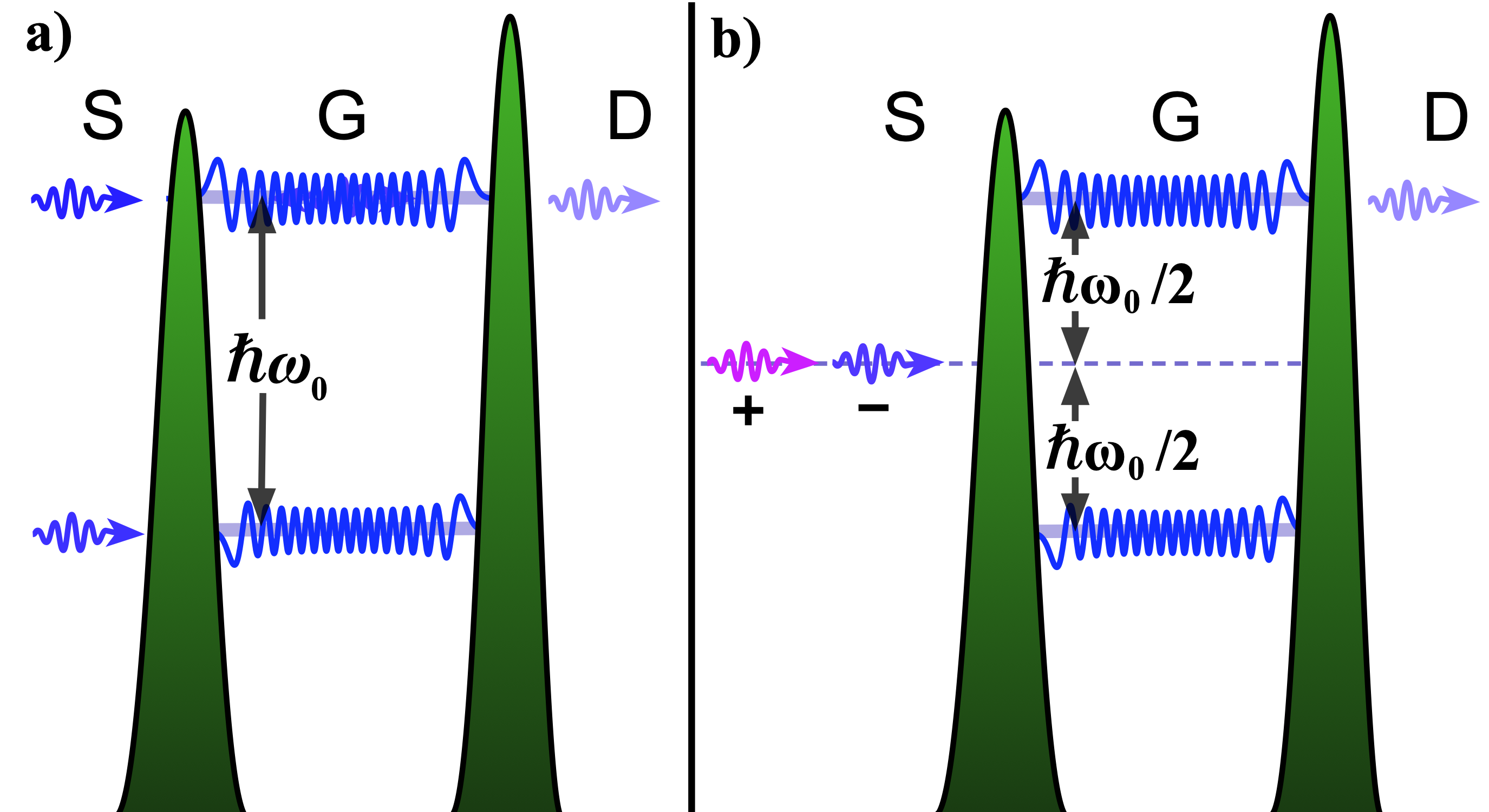}
\caption{\label{fig:NormalModes} In the absence of coupling particles travel from the source well to the drain via resonant tunneling of atoms having energy corresponding to the upper transistor level.  For the sake of illustration, incoming particles are depicted as arriving at the gate in free-space. Particles can also enter the gate via resonant tunneling for particles that carry energy corresponding to that of the lower transistor level. In each case the band of energies associated with gate particles is small, as is governed by the tunneling widths. All particles are reflected from the gate regardless of their energy except for the small portion of atoms that manage to transmit through to the drain from the upper level.  a) The reflection and transmission properties of the gate can be determined either in the original transistor mode basis, or b) in the normal-mode basis.  The latter is subsequently more convenient when analyzing the effects of mode coupling.  In the normal mode basis one considers modes ``arriving" at the gate having a mean energy and sidebands at $\pm \hbar\omega_0$.  Such modes can be associated with either symmetric ($+$) or anti-symmetric ($-$) combinations of the transistor modes.   }
\end{figure}

Tunneling of the upper transistor mode to the drain can be accommodated in terms of a transmission coefficient coupling the transistor mode to a mode of the drain. We write the amplitude transmission coefficient as $\sin(\kappa)$ so that:  
\begin{equation}
\begin{split}
	\left|\rm{SG}\right> =  \cos(\kappa)\left[\sin(\theta)\left|S_2\right>
	+\cos(\theta)\left |N+2\right> \right]\\
	+&\sin(\kappa)\left|D \right > .
\end{split}
\end{equation}
The transmission is linked to a decay constant $\Gamma_{\rm{T}}$ though a ``trial" rate $\gamma$:
\begin{equation}
	\Gamma_{\rm{T}}=\gamma \sin^2(\kappa).
\end{equation}
Typically we are interested in small transmissions, so
\begin{equation}
	\Gamma_{\rm{T}} \approxeq \gamma \kappa^2.
\end{equation}
While the decay constant arises from the conventional approach to open quantum systems (see for example \cite {scully1999}), both $\Gamma_{\rm{T}}$ and $\gamma$ are difficult to calculate for a complex potential, whereas the transmission coefficient is at least experimentally straightforward to adjust without substantially affecting $\gamma$ simply by changing the gate-drain barrier height or thickness.  
The introduction of tunneling implies that the transistor normal modes are no longer orthogonal:
\begin{equation}
	\left<-|+\right> =-\frac{1}{2}\kappa^2,
\end{equation}
in which the inner product is taken over a space that excludes the vacuum.  In other words, the normal modes are orthogonal in the absence of transmission, and are coupled when transmission to the vacuum is present. Said still another way, coupling to the vacuum introduces spontaneous emission between the normal modes, as suggested by re-writing the transistor Hamiltonian: 
\begin{equation}\label{Eq:TransistorHamiltonian}
\begin{split}
	\hat H_{\rm{T}}&=\left(N+2\right)\left(\hat b_+^\dagger \hat b_+ + \hat b_-^\dagger \hat b_- \right)\hbar\omega_0 \\
	& + \left( \hat b_+^{\dagger} \hat b_- + \hat b_-^\dagger \hat b_+\right)\frac{1}{2}\hbar\omega_0 .
\end{split}
\end{equation}
and noting the energy:
\begin{equation}
\begin{split}
	\left<E_{\rm{T}}\right>&=\left(N+2\right)\left(M_+ + M_- \right)\hbar\omega_0 \\
	& -  \frac{1}{2}\kappa^2\left( \sqrt{M_-(M_+ +1)}+\sqrt{M_+(M_- +1)}\right)\frac{1}{2}\hbar\omega_0 .
\end{split}
\end{equation}
We identify the second term in each of the above two expressions with spontaneous emission between the normal modes.  Spontaneous emission is a mechanism through which noise is introduced to the oscillator emission.

\section{The Many-Body Harmonic Oscillator}
Our operational ansatz is that the gate wavefunction forms as a displaced ground state of the harmonic potential. Mathematically this corresponds to coherent excitation and physically to dipole oscillation of the condensate.  We here clarify this ansatz by extending the usual simple harmonic oscillator treatment to a many-body framework.  The state that forms in the gate cannot in fact be a true displaced ground state because the number of gate levels is finite.  Nevertheless this initial assumption leads to a reasonable self-consistent result for the circuit  properties, and so we begin with a discussion of an ideal harmonic potential. 

 The physics of ultracold gases confined to harmonic potentials is a well-studied topic (see for example \cite{pethick2008}).  One is accustomed to Bose-condensation taking place in the many-body ground-state associated with the single particle ground state of the potential. In many-body physics the term ``coherent states" is typically meant to refer to eigenstates of the annihilation operators $\hat b_j$, of which Eq. (\ref{Eq:Symmetric}) and (\ref{Eq:AntiSymmetric}) are examples. These eigenstates do not have definite particle numbers.  In this section we work with a fixed particle number $M_{\rm{g}}$.  We argue that the whole of these $M_{\rm{g}}$ particles form a coherent state; to distinguish this from the conventional use of the term, we will refer to the $M_{\rm{g}}$-particle case as a ``massive coherent state" and refer to the more conventional eigenstate of the particle annihilation operator as a ``many-body coherent state".  To further avert likely confusion will continue to refer to the many-body particle operators labeled with a ``b" as \textit{creation} and \textit{annihilation} operators, and refer to the harmonic oscillator level-changing operators as \textit{ raising} and \textit{lowering} operators, or collectively as ladder operators, labeled with the letter ``a". Thus we can write the definition of a many-body coherent state $\left|\beta\right>$ as:
 \begin{equation}
 	\hat b_j \left| \beta \right> = \beta \left| \beta \right >.
 \end{equation}
 We focus specifically on the massive coherent state for the remainder of this section. 

One is familiar with the actions of the ladder operator on Fock states $\left|n\right>$ of the simple (single particle) harmonic oscillator:
\begin{eqnarray}
	\hat a\left|n\right> = \sqrt{n}\left|n-1\right> \\
	\hat a^\dagger\left|n\right>=\sqrt{n+1}\left | n+1\right>.
\end{eqnarray}
The wavefunctions corresponding to the Fock states are given by Hermite-Gaussians:
\begin{equation}\label{Eq:HermiteGaussians}
	\psi_n(x)=\frac{1}{\sqrt{2^n n!}}\left(\frac{m\omega_0}{\pi\hbar}\right )^{1/4} e^{-\frac{m\omega_0}{2\hbar} x^2} H_n \left( \sqrt{\frac{m\omega_0}{\hbar}} x \right),
\end{equation}
in which $H_n(x)$ is the $\rm{n}^{th}$-order Hermite polynomial, $\omega_0$ is the resonant frequency of the harmonic oscillator, and $m$ is the particle mass.
The ladder operators obey the commutation relation:
\begin{equation}
	[\hat a, \hat a^\dagger]=1,
\end{equation}
and Hamiltonian of the simple harmonic oscillator is:
\begin{equation}\label{Eq:HamiltonianSHO}
	\hat H_{\rm{SHO}}=\left(\hat a^\dagger \hat a+\frac{1}{2}\right)\hbar \omega_0.
\end{equation}
The canonical coherent state is defined as an eigenstate of the lowering operator:
\begin{equation}	\hat a \left|\alpha \right>=\alpha \left|\alpha \right>.
\end{equation}
Such a state can be generated from the ground state using a displacement operator:
\begin{equation}\label{Eq:SingleParticleCS}
	\hat D(\alpha)\left |0 \right> = \left|\alpha \right>,
\end{equation}
in which
\begin{equation}
	\hat D(\alpha) \equiv \exp \left(\alpha \hat a^\dagger-\alpha^* \hat a \right).
\end{equation}
A coherent state can be expressed as a superposition of Fock states:
\begin{equation}
	\left|\alpha \right> = e^{-|\alpha|^2/2}\sum_{n=0}^\infty\frac{\alpha^n}{\sqrt{n!}}\left|n\right>.
\end{equation}
The expectation value of the energy can be obtained from the Hamiltonian Eq. (\ref{Eq:HamiltonianSHO})
\begin{equation}
	\left<E\right > = \left(\left|\alpha\right|^2+\frac{1}{2}\right)\hbar\omega_0.
\end{equation}

Our ansatz utilizes coherent states as the single-particle basis in which to build a many-body state. We use the following notation to designate a massive coherent state:
\begin{equation}\label{Eq:MassiveCoherentState}
	\left|\alpha\right>\rightarrow  \left |\alpha;M_{\rm{g}}\right>,
\end{equation}
that is, a many-body displaced ground state that is occupied by $M_{\rm{g}}$ particles. The expectation value of the energy of this state is given directly from the single particle case:
\begin{equation}\label{Eq:MassiveCSEnergy}
	\left<E_\alpha\right > =M_{\rm{g}} \left(\left|\alpha\right|^2+\frac{1}{2}\right)\hbar\omega_0.
\end{equation}

An alternative approach to our ansatz recognizes that the dynamics of the center-of-mass of the many-body oscillator can be written independently from the dynamics of the relative coordinates of the individual particles.  The center-of-mass wavefunctions will be Hermite-Gaussians as in Eq. (\ref{Eq:HermiteGaussians}), except with the mass replaced by $M_{\rm{g}}\times m$.  In particular, though it has greater mass, the center of mass energy spectrum is identical to the single particle case, $E_n = (n+1/2)\hbar \omega_0$ because it also experiences a force constant that is $M_{\rm{g}}$ times larger than that of the single particle.  One again arrives at Eq. (\ref{Eq:MassiveCoherentState}) with through the assumption that the center-of-mass wavefunction is a coherent state, while the relative coordinate wavefunctions are all in their ground state.  One reproduces Eq.  (\ref{Eq:MassiveCSEnergy}) through appropriate scaling of the oscillation amplitude $\alpha$.

\section{Gate Interaction Energy}

We are now equipped to return to the analysis of the model for the gate well. One might wonder how our ansatz,  the presence of oscillating condensate, can be associated with the gain needed to maintain oscillation. Recognize that particles which enter an initially empty gate find themselves in a region of population inversion —at the top of a ladder of levels leading to a ground state. In his wonderfully pedagogical article, Glauber \cite{GLAUBER.1986} describes how such an \textit{inverted} ladder system coupled to a reservoir leads to gain rather than damping.  Indeed, in his  ``inverted" harmonic oscillator, he establishes that an initial small coherent oscillation amplitude grows exponentially in time, extracting energy from the reservoir rather than dissipating energy into it. Glauber uses this harmonic oscillator-plus-reservoir as a model for classical gain in an electromagnetics context. Whereas his Gedanken oscillator draws power from the reservoir at an exponentially increasing rate, in our case power is provided by the chemical potential of particles in the source well, which rate is necessarily limited. That limited rate leads to saturation and consequently a constant oscillation amplitude. 

Our objective in this section is to determine the interaction between the oscillating condensate in the gate and the transistor modes.  The transistor mode spacing is degenerate with the oscillation frequency of the condensate, and this leads to phase-coherence between the transistor states, which is what led us to define the normal modes basis earlier. Determining the coupling interaction energy sets the stage for understanding how the interaction can lead to a self-consistent oscillation amplitude $\alpha$ together with a particle current flowing to the drain.  In the following section we show that the drain current leads to cooling of the gate particles —a conclusion that is consistent with the kinetic theory.  Even for modest amplitudes of oscillation we can expect $|\alpha|\gtrsim k_{\rm{B}}T_{\rm{G}}/\hbar\omega_0$, in which case the uncertainty in particle energy associated with a coherent state is larger than the spread of energies associated with the gate temperature.  Thus for the purposes of this section, we assume that the gate particles all occupy a common state, approximately the canonical massive coherent state Eq. (\ref{Eq:MassiveCoherentState}) . 

The many-body interaction Hamiltonian, Eq. (\ref{Eq:ManyBodyHamiltonian}), can be separated into three contributions, 
\begin{equation}\label{Eq:ThreeInteractions}
	\hat H_{\rm{GI}}=\hat H_{\rm{GG}}+\hat H_{\rm{MM}}+\hat H_{\rm{GT}}.
\end{equation}
The first, referring to the interaction of gate particles, involves particles operators corresponding to the gate bound states:
\begin{equation}\label{Eq:GateSelfInteraction}
\hat H_{\rm{GG}}=\frac{\eta}{2}\sum_{ijkl=0}^{N} U_{ijkl} \hat b_{i}^{\dagger}\hat b_{j}^{\dagger}\hat b_{k}\hat b_{l} ,
\end{equation}
Other than the finite set of levels involved, this is the usual interaction Hamiltonian for the many-body harmonic oscillator.  For a condensate in the ground state, the contribution $H_{0000}$ corresponds to the mean-field energy that shifts the chemical potential.  This mean field energy remains the same for the displaced ground state, at least when the mean field is small compared to the level spacing.  The value of the mean field is determined self-consistently in a later section.  Again considering a condensate in the ground state, contributions such as $H_{0j0j}$ correspond to mean-field shifts of energy levels.  This does not impact the level spacing, and so qualitatively does not change the physics of our system.  The same is true when the ground state is displaced. Other self-energy terms lead to the Bogolyubov excitation spectrum.  We ignore these in detail but they are implicitly treated as a source of noise and (gate-internal) dissipation associated with our oscillator. 

The second term of Eq. (\ref{Eq:ThreeInteractions}) accounts for mass exchange between the gate and transistor modes.  Accordingly, the summation includes terms that have three particle operators belonging to either the gate or to transistor modes, and one from the other. The mass exchange describes the approach to chemical potential equilibrium between the transistor and gate modes.  For our purposes it will suffice to simply assume chemical equilibrium self-consistently. 

The last term, $\hat H_{\rm{GT}}$ is our primary interest. This coupling between gate and transistor modes is mediated by phonon interactions with the condensate. 

While the canonical coherent state involves all oscillator modes, our transistor model assumes that a finite set of levels can participate. We introduce the \textit{truncated} coherent state, $ \left| {\tilde \alpha } \right\rangle$. In the (single particle) Fock basis:
\begin{equation}\label{Eq:TruncatedCoherentState}
	 \left| {\tilde\alpha} \right\rangle=\sqrt{C_N(\tilde \alpha)} {e^{ - {{{{\left| \tilde \alpha  \right|}^2}} \mathord{\left/
 {\vphantom {{{{\left| \tilde \alpha  \right|}^2}} 2}} \right.
 \kern-\nulldelimiterspace} 2}}}\sum\limits_{n = 0}^N  {\frac{{{\tilde \alpha ^n}}}{{\sqrt {n!} }}\left| {n} \right\rangle } ,
\end{equation}
where $C_N(\tilde\alpha)$ is a normalization factor:
\begin{equation}
C_N^{ - 1}\left( \tilde \alpha  \right) = {e^{ - {{\left| \tilde \alpha  \right|}^2}}}\sum\limits_{n = 0}^N {\frac{{{|\tilde \alpha |^{2n}}}}{{n!}}}.
\end{equation}
For notational simplicity we will drop the tilde from the state amplitude $\alpha$.  Eq. (\ref{Eq:TruncatedCoherentState}) describes a particle undergoing dipole oscillation with complex amplitude $\alpha=|\alpha|e^{\phi_{\rm{g}}}$ and frequency $\omega_0$. 

\begin{figure}
\includegraphics[width=\columnwidth]{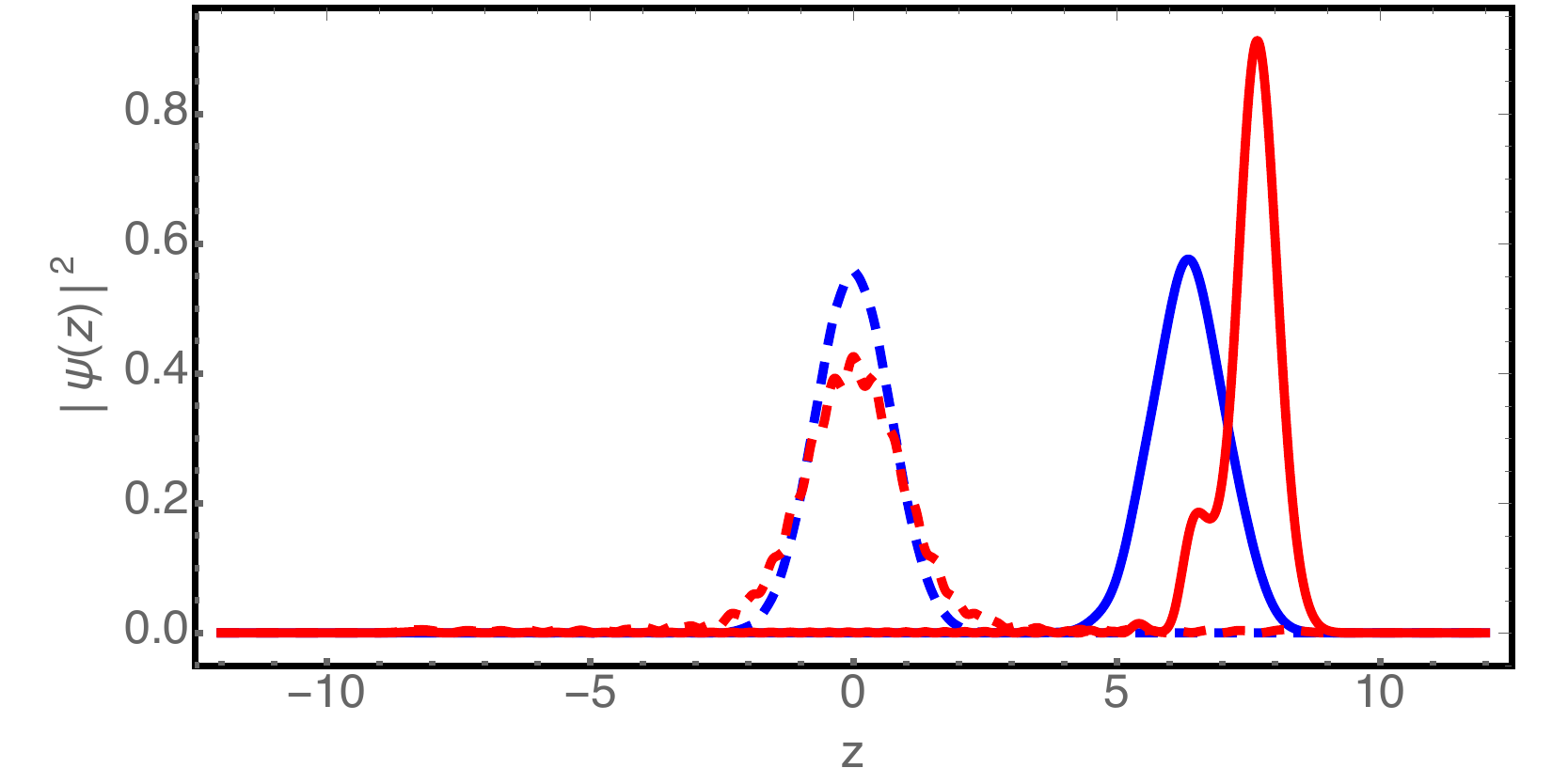}
\caption{\label{fig:TruncatedStates} Probability density for truncated coherent states corresponding to a harmonic oscillator having $N=36$; the z-axis is normalized to units of $\left(\hbar/m \omega_0\right)^{1/2} $.  The solid lines correspond to the state at the classical turn-around time while the dashed lines correspond to the moment its position is at the middle of the potential. The blue curves correspond to an amplitude of $|\alpha|^2/N=0.56$ while the red curves correspond to an amplitude of $|\alpha|^2/N=0.84$.}
\end{figure}

A harmonic oscillator having a finite set of levels cannot be associated with a set of ladder operators in the usual way; however, the truncated wavefunctions reasonably approximate those of the canonical coherent state provided $|\alpha|\lesssim\sqrt{N}$ and in this same limit $C_N(\alpha)\simeq 1$. At the same time we will find that the solution to self-consistent dynamics indicates states whose oscillation amplitudes is placed near the border of the inequalities.  Fig. \ref{fig:TruncatedStates} shows truncated states corresponding to two different oscillation amplitudes for $N=36$.  The smaller amplitude closely approximates the shape of a coherent state while it is notable that the larger amplitude exhibits squeezing. We note that the squeezing is enhanced as $N$ becomes larger and as the amplitude reaches its maximum value.  We additionally comment that mean field effects will alter the mode's shape at large gate chemical potentials.

Having laid the foundation in the previous section, a premise of this work is that phonon absorption or emission by the transistor states corresponds to phonon emission or absorption primarily by the \textit{whole} of the condensate. As indicated earlier, the center-of-mass energy eigenstates have the same energy spacing $\hbar \omega_0$ as the individual particles; though phonon energy is quantized, energy gained or lost is shared among the particles. The result is a change of oscillation amplitude corresponding to an energy change that is a fraction of $\hbar\omega_0$ per particle.  
The relevant phonon interaction Hamiltonian is derived from Eq. (\ref{Eq:ManyBodyHamiltonian}) in conjunction with our assumed gate statevector Eq. (\ref{Eq:TruncatedCoherentState}):
\begin{equation}\label{Eq:InteractionHamiltonian}
\begin{split}
	\hat H_{\rm{GT}}  &= 2\eta{C_N} e^{ - {{ |\alpha |^2}}}\sum\limits_{n = 1}^N {U_n}\frac{{\alpha ^{2n - 1}}}{\sqrt {n!(n - 1)!} } \\ 
	&\times \hat b_{n-1}^\dagger \hat b_{\rm{SD}}^\dag \hat b_{n}{\hat b_{\rm{SG}}}  + {\rm{h.c.}}  
\end{split}
\end{equation} 
where $\rm h.c.$ stands for \textit{Hermitian conjugate} and the overlap factors are:
\begin{equation}\label{Eq:OverlapIntegral}
	U_n= \int {\psi _{n-1}^ *(x)\psi _{n}(x) \psi _{\rm{SD}}^ *(x)\psi _{\rm{SG}}(x) dx}.
\end{equation}
The supposition that the gate is approximately in a displaced ground state corresponding to an oscillating dipole with amplitude $\alpha$ means that the operators $\hat b_{n-1}^\dagger b_n$ are independent of mode number and can be replaced with a c-number:
\begin{equation}\label{Eq:cnumber}
	\hat b_{n-1}^\dagger b_n\rightarrow M_{\rm{g}}; \left\{n\leq N \right \}.
\end{equation}

The primary focus of this work is on the system-as-oscillator, establishing a self-consistent result stemming from the assumption that the gate exhibits dipole oscillation in the form of a (truncated and massive) coherent state. With this in mind we invoke the assumption that the stationary solution for the gate and transistor states can be approximated as a direct product of pure coherent states:
\begin{equation}\label{Eq:GateStateDirectProduct}
	| G \rangle \approx {|\alpha ;M_{\rm{g}} \rangle  |+;M_+ \rangle|-;M_- } \rangle .
\end{equation}
The direct product approximation enables a straightforward calculation of the interaction energy. Before continuing to the calculation, however, it is perhaps useful to point out the key underlying assumptions that have brought us this far, and reflect on aspects that could lead to their breakdown: Given the replacement of particle operators with a c-number, Eq. (\ref{Eq:cnumber}), we recognize that the gate state $|\alpha ;M_{\rm{g}} \rangle $ is itself assumed to be a direct product of the individual particle states of the gate. In particular our assumed massive coherent gate state ansatz is a special case of a direct product state.  While the many-body ground state (i.e. the Bose-Einstein condensate) of a harmonic oscillator potential is indeed a direct product state, we do not establish here that the same is necessarily true in our dynamical case.  Interactions among the gate particles, embedded in Eq. (\ref{Eq:GateSelfInteraction}), which can be significant compared with the interaction energy calculated below, can possibly lead to an entangled gate state.  Further, we can expect that interactions of the transistor modes with the reservoir modes of the source well lead to mixed states.  Under usual circumstances such mixing would lead to damping of coherence between the transistor modes.  Left to its own, the dipole oscillation of the condensate will decay, reflecting finite quality factor ($Q$) and other factors.  If oscillation is to take place, system gain must be present to overcome damping and drive coherence.  We return to this issue in a later section discussing oscillation threshold.  In the presence of oscillation, the underlying dissipation effects, both positive and negative, will manifest as noise in the oscillation phase and amplitude.  Finally, interaction between the transistor modes and the gate can cause the two to be entangled, possibly leading to a squeezed oscillation output as alluded to above. The topics of noise, entanglement, and other interesting and admittedly important aspects of this atomtronic system are left to future works.

\begin{figure}
\includegraphics[width=\columnwidth]{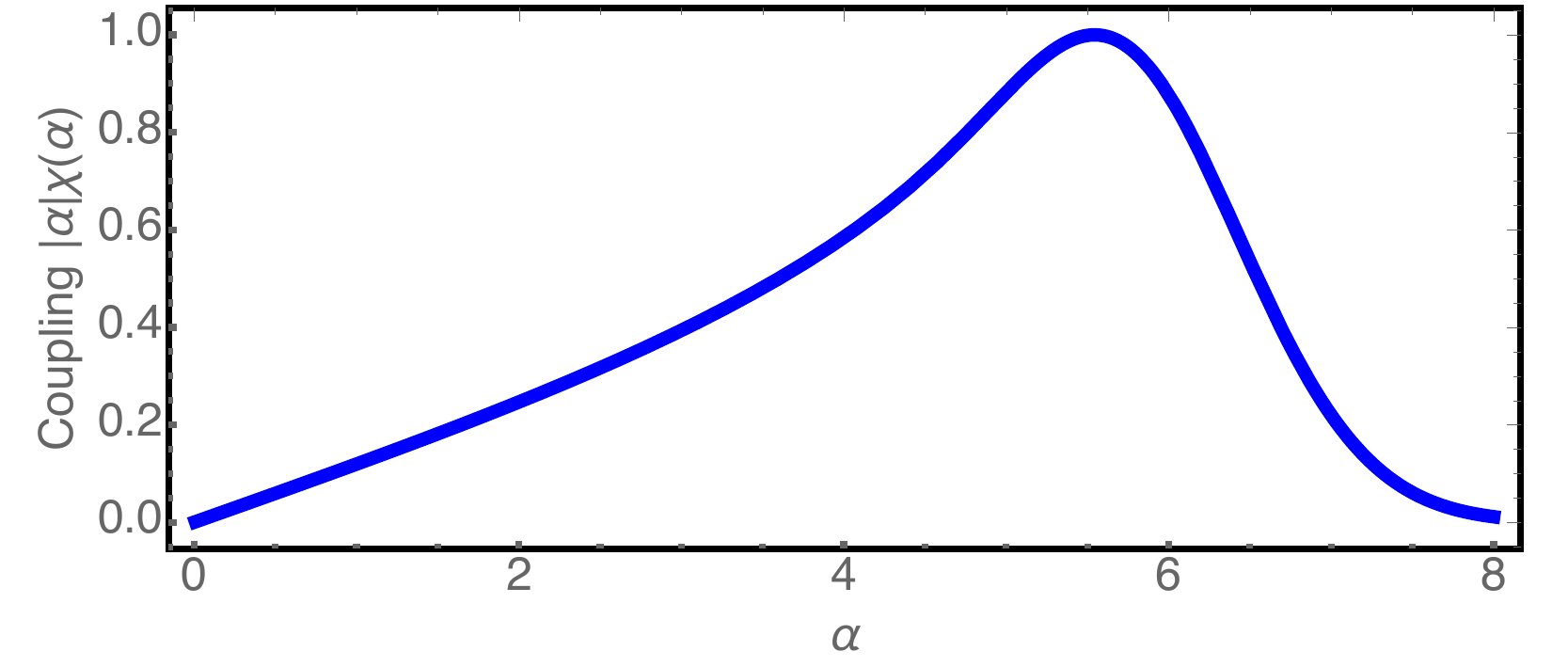}
\caption{\label{fig:CouplingEnergy} Coupling factor $|\alpha|\chi (\alpha)$ relative to its maximum value for a truncated harmonic oscillator having $N=36$ levels (plus the ground state).  Notice that the peak occurs near $|\alpha|=\sqrt{N}=6$. }
\end{figure}

The interaction energy can be found by substitution of Eq.s (\ref{Eq:SymmetryOperators}), (\ref{Eq:OverlapIntegral}),  and (\ref{Eq:GateStateDirectProduct}) into Eq. (\ref{Eq:InteractionHamiltonian}):
\begin{equation}\label{Eq:GateInteraction}
\begin{split}
	\left< E_{\rm{GT}}\right> & = \chi\left( \alpha  \right)  |\alpha| M_{\rm{g}} 
	 \left[\cos \left( \phi \right)\left( M_+ - M_- \right)  - i\frac{1}{2}\kappa^2\sin \left( \phi  \right) \right. \\
	& \times \left. \left( \sqrt{M_+(M_-+1)} - \sqrt{M_-(M_++1)} \right) \right].
\end{split}	
\end{equation}
Here the \textit{coupling} factor is:
\begin{equation}\label{Eq:chi}
\chi \left( \alpha  \right) \equiv 4\eta C_N\left( \alpha  \right)e^{ - \left| \alpha  \right|^2}\sum\limits_{n=1}^N \frac{|\alpha| ^{2(n - 1)}}{\sqrt {n! (n - 1)!} } |U_n|.
\end{equation}
The phase $\phi$ is the difference between the dipole oscillation phase $\phi_{\rm{g}}$ and the phase associated with the symmetric transistor (normal-mode) state: $\phi
\equiv\phi_{\rm{g}}-(\phi_{\rm{SD}}-\phi_{\rm{SG}})$.  Fig. \ref{fig:CouplingEnergy} presents a plot of the coupling factor $|\alpha| \chi (\alpha)$ as a function of the oscillation amplitude for a gate oscillator having $N+1=36+1$ levels.  For small amplitudes the coupling factor is linear in $\alpha$ while it peaks and then falls to zero at larger amplitude.  The peak occurs a bit below $\sqrt{N}=6$.

This concludes the development of the many-body treatment of the transistor gate.  The gate interaction energy, Eq. (\ref{Eq:GateInteraction}), along with the coupling constant Eq. (\ref{Eq:chi}), is the central mathematical result of this section. 

Two aspects are of particular note. The first is that the interaction energy is zero if the number of particles in the symmetric and anti-symmetric states are equal.  The second is that the energy is dependent on the relative phase between the oscillating dipole and the transistor modes.  The transistor modes themselves are coupled to a reservoir particles in the source well characterized by a temperature and chemical potential.  There is nothing \textit{a priori} that determines the occupation or phase of these modes.  

\section{A Classical Equivalent Circuit}

While the key physics of transistor behavior is implicit in Eq. (\ref{Eq:GateInteraction}), understanding transistor action and the conditions under which oscillation can occur are vastly easier to comprehend by transitioning to continuous variables and a classical picture involving chemical potential differences $\mu$ and particle currents $I$.  For example, we write the particle number in the gate in terms of a capacitance:
\begin{equation}
	M_{\rm{g}} = \mu_{\rm{g}} C_{\rm{g}}.
\end{equation}
We have defined the capacitance as the electronic dual, for which the unit is the Farad, i.e. $\rm{C}^2/J$.  The atomtronics capacitance here is given in units of $(\rm {particle})^2/J$  We can likewise write the flow of particles per second into the drain as a current:
\begin{equation}
	I_{\rm{d}}=\frac{1}{2}\Gamma_{\rm{T}} \left(M_+ +M_-\right).
\end{equation}
\begin{figure}
\includegraphics[width=\columnwidth]{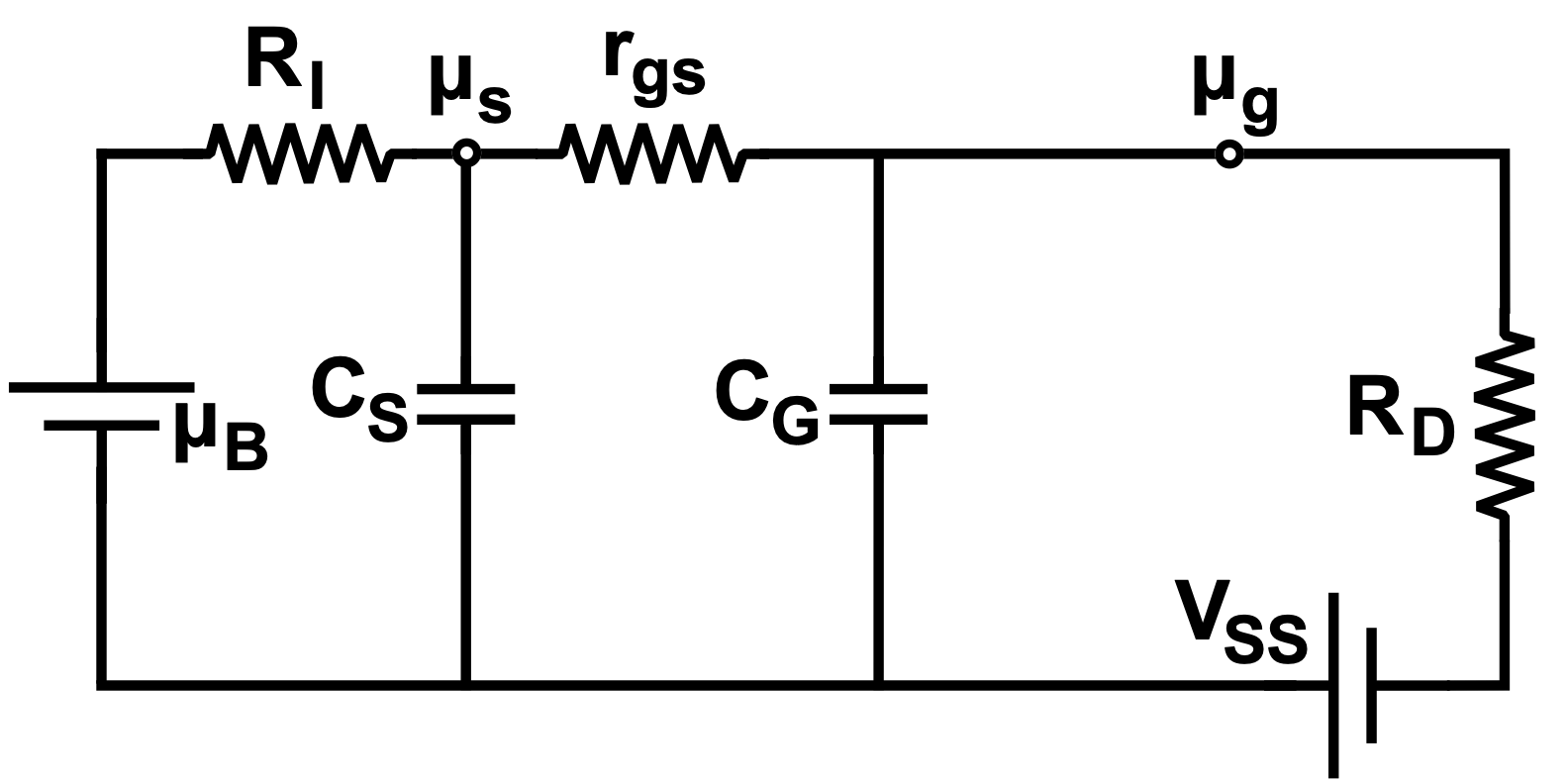}
\caption{\label{fig:Circuit} The atomtronic transistor oscillator ``DC" equivalent circuit consists of a battery having an internal resistance along with a bias potential which physically corresponds to the bias shown in Fig. \ref{fig:BiasedTransistor}.  The transistor itself is characterized by a negative trans-resistance and a gate capacitance that sets the number of particles in the gate.  The drain resistance reflects the fact that the output current carries energy that can be used to do work (in this case, ``heating'' the vacuum).}
\end{figure}

Within the classical picture we can make substantial use of the DC equivalent circuit that is shown in Fig. \ref{fig:Circuit}. The circuit is ``powered" by the chemical potential provided by the source particles, represented as a battery potential $\mu_{\rm{B}}$ and having an added bias potential shown as $V_{\rm{SS}}$.  Any battery is \textit{fundamentally} associated with an internal resistance $R_{\rm{I}}$. The internal resistance associated with an electrical battery is always positive and leads to heating of the battery as current flows. An atomtronic battery is associated with an internal resistance that can be either positive \textit{or} negative, depending, respectively, on whether it supplies condensed or thermal atoms.  In the latter case, the resistance causes cooling of the battery, a manifestation of forced-evaporation techniques used to produce Bose-condensates, for example. As is the case in electronics, a holistic circuit design would include an appropriately chosen resistance in series with the battery, as in Fig. \ref{fig:ColpittsFET}; here we lump a possible added series resistance into the battery internal resistance.  

The drain resistance $R_{\rm{D}}$ is positive and accounts for the energy lost to the vacuum as particle current flows from the gate to the drain. The source capacitance $C_{\rm{S}}$ accounts for the chemical potential developed in the source as particles flow from the battery.  Likewise $C_{\rm{G}}$ accounts for the chemical potential developed in the gate as particles accumulate there in a (displaced) condensate.  The gate-source resistance, $r_{\rm{gs}}$, represents the change in potential energy experienced by particles flowing from the source to the gate. Note in particular that we refer to the source well potential $\mu_{\rm{s}}$ \textit{after} current has flowed from the battery through the internal resistance. 

In absence of the coupling we suppose that the battery provides a population to the transistor modes in the ratio determined by a Boltzmann factor: 
\begin{equation}
	\frac{M_{N+2}}{M_{N+1}}=\exp\left(-\frac{\hbar\omega_0}{k_{\rm{B}} T_{\rm{B}}}\right).
\end{equation}
The upper state population leads to a corresponding bias current in the presence of tunneling that can be written:
\begin{equation}\label{Eq:BiasCurrent}
\begin{split}
	I_0 &= \mu_{\rm{B}} C_S \Gamma_{\rm{T}}   \exp\left(-\frac{\hbar\omega_0}{k_{\rm{B}} T_{\rm{B}}}\right) \\
	&\equiv I_{\rm{d}} \leftarrow  \textnormal{} \left(\left<E_\gamma\right>=0\right),
\end{split}
\end{equation}
in which $I_0$ is contributed to equally by symmetric and anti-symmetric modes.  Eq. (\ref{Eq:BiasCurrent}) supposes the bias current is small so that the drop across the internal resistance is negligible. Given fixed battery temperature and chemical potential as well as barrier heights, the bias current will exponentially depend on the applied bias potential, $V_{SS}$.  

Eq. (\ref{Eq:GateInteraction}) indicates that the coupling energy is zero when the normal mode populations are equal.  This will be the case, for example, when the gate-drain barrier is infinite such that the gate and source come into thermal equilibrium.  We  will show below however, that another, self-consistent solution, exists in which only one of the symmetric or anti-symmetric modes arises in the transistor and a significant drain current, either $I_+$ or $I_-$, will flow. In other words, either $I_{\rm{d}}=I_+$, $I_-=0$ or $I_{\rm{d}}=I_-$, $I_+=0$. While both normal modes corresponds to having particles equally likely found in the lower and upper transistor states, particles can escape to the drain only in the upper of the two.  On the other hand, particles entering from the battery occupy the lower and upper states with a ratio that depends on the temperature, such that the probability that the particle entered in the lower state is:
\begin{equation}
	\varrho=1-\frac{1}{2} \exp\left(-\frac{\hbar\omega_0}{k_{\rm{B}} T_{\rm{B}}}\right).
\end{equation}

Essentially all of the significant physics that follows arises from a pair of conclusions related to the fact that current flow into the drain corresponds to the removal of heat from the system:
\begin{equation}
	P_{\rm{d}}=\varrho\hbar\omega_0 I_{\rm{d}}.
\end{equation}
The first conclusion is that the removal of heat is corresponds to a \textit{negative} gate-drain resistance $r_{\rm{gs}}$:
\begin{equation}
\begin{split}
		P_{\rm{gs}} &= -|r_{\rm{gs}}|I_{\rm{d}}^2 \Rightarrow \\
		r_{\rm{gs}}&=-\varrho\frac{\hbar\omega_0}{I_{\rm{d}}}.
\end{split}
\end{equation}
The removal of heat indicates that the gate temperature will be lower than the battery temperature. The source  potential is of course lower than the battery potential by the resistive drop:
\begin{equation}
	\mu_{\rm{s}}=\mu_{\rm{B}} - I_{\rm{d}}R_{\rm{I}}.
\end{equation}
The second conclusion is that the gate potential is in fact higher than the source potential, 
\begin{equation}
	\mu_{\rm{g}}=\mu_{\rm{s}}+\varrho \hbar \omega_0 .
\end{equation}
Thus the current flows into the gate \textit{against} the potential drop.  These two conclusions echo those of the kinetic treatment.

Most importantly, cooling of the gate supports the assumption that the gate particles are condensed.  Moreover, incoming particles from the battery are very cold in the first place, yet to form a condensate in the ground state of the gate requires that the particles first loose a great deal of kinetic energy.  Forming a condensate that itself has kinetic energy requires substantially less cooling that a ground-state condensate would.

\section{Current Gain and Steady-State}
The mechanism by which transistor action controls the drain current is through the coupling energy, Eq. (\ref{Eq:GateInteraction}). To be specific we suppose that the transistor levels are populated by the antisymmetric mode, so that:
\begin{equation}	
	\left<E_\gamma\right> \approxeq -\chi(\alpha)|\alpha| C_{\rm{G}} \mu_{\rm{g}} M_-\cos(\phi),
\end{equation}
where we have dropped the second term since the transmission $\kappa^2$ is small compared with unity.  With respect to the oscillation amplitude, the coupling energy has a maximum value of $\chi(\alpha)|\alpha|$ as illustrated in Fig.\ref{fig:CouplingEnergy}. To a reasonable approximation:
\begin{equation}
	 \max\left[ \chi(\alpha)|\alpha| \right]\approxeq \chi_0\sqrt{N},
\end{equation}
where $\chi_0 \equiv \chi(0)$.  We can thus write for this case that the maximum coupling energy per particle that contributes to the drain current as:
\begin{equation}
	V(\phi)=-V_0 \cos (\phi),
\end{equation}
in which,
\begin{equation}\label{Eq:PotentialEnergyAmplitude}
		V_0 \approxeq \chi_0\sqrt{N} C_{\rm{G}} \mu_{\rm{g}}.
\end{equation}

We recognize the coupling to be a \textit{phase-dependent} potential energy.  Herein lies the origin of the current gain of the transistor:  In absence of coupling, only particles having energy within a narrow band around $E_{\rm{T}} = (N+2)\hbar \omega_0$ can contribute to the drain current (see Fig. \ref{fig:NormalModes}).  In the presence of coupling, however, particles corresponding to normal modes nominally having too much or too little energy can nevertheless contribute to the current provided they are associated with the appropriate phase.  (Here we are speaking loosely, referring to a ``particle's phase" where more appropriately speaking, it is the phase of the interference of the two wavefunctions corresponding to upper and lower transistor states relative to the phase of the oscillating truncated coherent state).  In other words, such particles can borrow energy from, or lend energy to, the interaction energy.  Once again to be specific we consider particles associated with the antisymmetric modes. Particles having an energy $E_{\rm{T}}+\delta E$ can participate in the drain current provided they are associated with a phase:
\begin{equation}\label{Eq:MatchedPhase}
	\phi = \arccos\left(\frac{\delta E}{V_0}\right).
\end{equation}
Assuming the density of states is constant within the range $E_{\rm{T}} \pm V_0$, the total drain current is given by the ratio of the coupling energy to the tunneling energy:
\begin{equation}\label{Eq:DrainCurrentfromCoupling}
	I_{\rm{d}}=\frac{1}{1-\varrho}\frac{V_0}{\hbar \Gamma_{\rm{T}}	}I_0=\mu_{\rm{B}} C_{\rm{S}}\frac{V_0}{\hbar},
\end{equation}
where the prefactor takes into account the Boltzmann factor for the relative number of atoms in the lower compared to the upper transistor state. We thus see that the presence of coupling induces a current gain:
\begin{equation}
	\beta=\frac{I_{\rm{d}}}{I_0}=\frac{1}{1-\varrho}\frac{V_0}{\hbar \Gamma_{\rm{T}}}.
\end{equation}

\begin{figure}\includegraphics[width=\columnwidth]{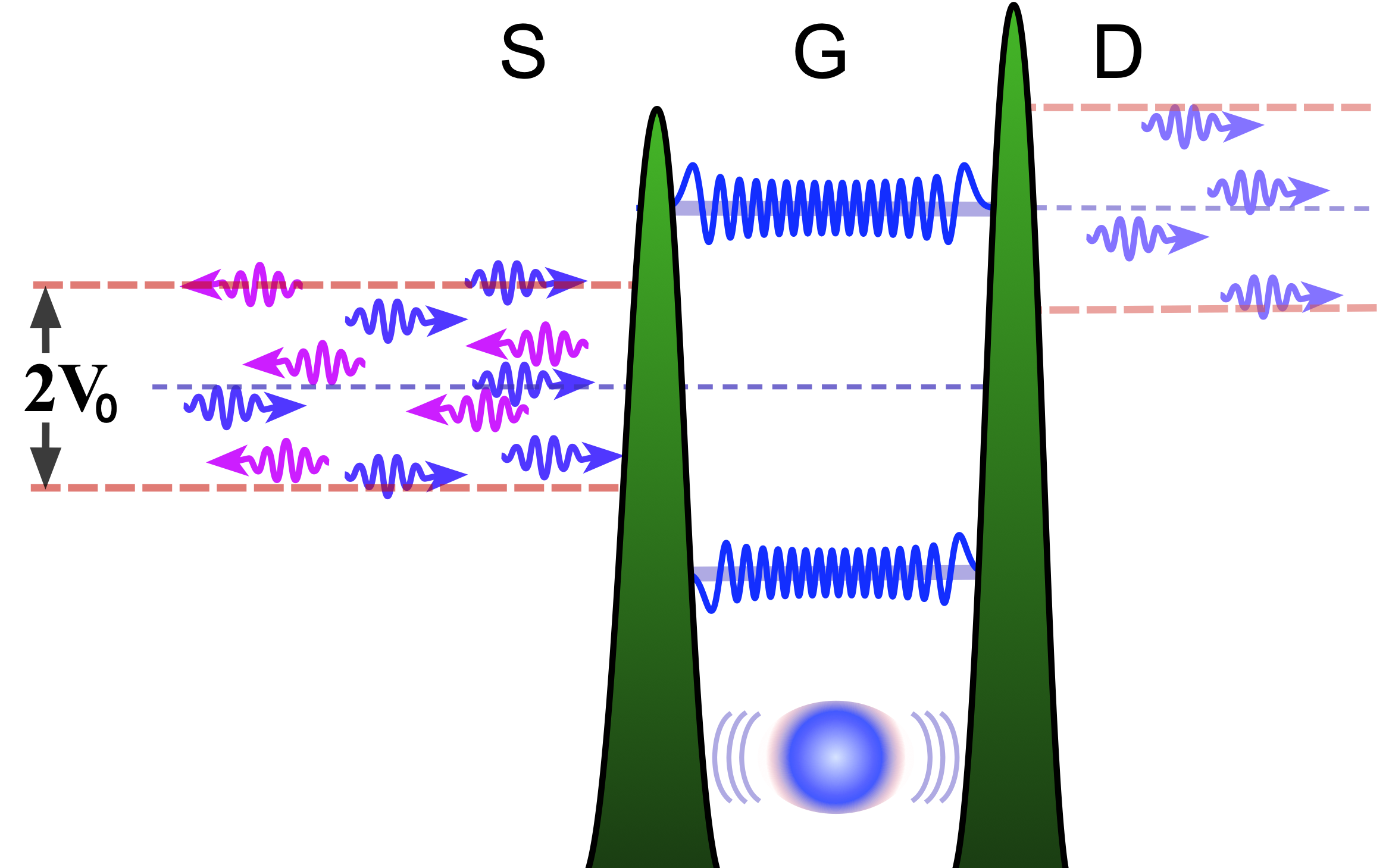}
\caption{\label{fig:CurrentGain} In the presence of coupling to the oscillating condensate, only one of the two normal modes enters the gate by resonant tunneling (here, the antisymmetric modes shown in blue).  The other is reflected (the symmetric modes, shown as magenta). The interaction energy enables large range of incident particle energies to enter into the gate in comparison to the interaction-free case. The drain current is thus also much larger.   }
\end{figure}

We are now in a position to understand the manner in which the single-normal-mode enters into a self-consistent solution.  The complement phase $\phi(\delta E)+\pi$ of the matched phase of Eq. (\ref{Eq:MatchedPhase}) is associated with a symmetric mode, the energy of which is non-resonant with the transistor modes. The net impact of the interaction energy is illustrated in Fig. \ref{fig:CurrentGain}.   The exception is the small band of energies around $E_{\rm{T}}$ with $\phi=\pi$: this small band can be thought of as responsible for spontaneous emission and fluctuations in the drain current. Though important to fully characterize circuit behavior we neglect their contribution in this work.  

The open-loop properties of electronic field-effect transistors are typically characterized in terms of a transconductance:
\begin{equation}
	I_{\rm{d}}=g_m \mu_{\rm{g}}.
\end{equation}
Combining Eqs. (\ref{Eq:PotentialEnergyAmplitude}) and (\ref{Eq:DrainCurrentfromCoupling}), evidently
\begin{equation}
	g_m=\chi_0\sqrt{N} C_{\rm{G}} C_{\rm{S}} \mu_{\rm{B}} / \hbar.
\end{equation}
Our oscillator circuit operates in closed loop, such that the gate potential depends on the drain current,
\begin{equation}
	\mu_{\rm{g}}=\mu-I_{\rm{d}} R_{\rm{I}},
\end{equation}
in which the driving potential $\mu$ is defined as:
\begin{equation}
	\mu\equiv \mu_{\rm{B}}+\varrho \hbar \omega_0.
\end{equation}
Therefore the steady-state circuit drain current is:
\begin{equation}\label{Eq:DrainCurrent}
	I_{\rm{dss}} = \frac{g_m}{1+g_m R_{\rm{I}}}\mu.
\end{equation}

We note that in steady-state, the coupling energy, Eq. (\ref{Eq:PotentialEnergyAmplitude}), averages to zero.  Moreover and equivalently, the transistor normal mode oscillation is $90^\circ $ out of phase with the condensate dipole oscillation.  We would expect as much for a driven or a forced oscillator operating in steady-state.  

In steady-state, power is dissipated by the battery internal resistor, leading to heating of its reservoir of particles:
\begin{equation}
	P_{\rm{b}}=I_{\rm{dss}}^2 R_{\rm{I}}=\left(\frac{g_m \mu}{1+g_m R_{\rm{I}}}\right )^2 R_{\rm{I}}.
\end{equation}
At the same time, the drain cooling causes cooling of the gate:
\begin{equation}
	P_{\rm{g}}=-\varrho\hbar\omega_0 I_{\rm{dss}} =- \frac{g_m \mu }{1+g_m R_{\rm{I}}}\varrho\hbar\omega_0.
\end{equation}

Our discussion thus far treats the gate as ideally harmonic, meaning equally spaced energy levels.  In practice, anharmonicity of the gate potential will cause decoherence among the oscillator levels.  Resonant circuits are typically characterized by a quality factor:
\begin{equation}
	Q\equiv \frac{\omega_0}{\Gamma_{\rm{osc}}},
\end{equation}
in which $\Gamma_{\rm{osc}}$ characterizes the linewidth, or inverse coherence time of the oscillator. The finite quality fact implies the kinetic energy of the oscillator is converted into heat at a rate:
\begin{equation}
	P_{\rm{osc}}=\Gamma_{\rm{osc}} \mu_{\rm{g}} C_{\rm{G}}|\alpha|^2\hbar\omega_0  .
\end{equation}
The dissipated power can be incorporated in the equivalent circuit by adding a resistance in series with the (negative) gate source resistance.  Cooling of the gate through $r_{\rm{gs}}$ will compensate for the dissipated oscillator energy provided:
\begin{equation}
	\varrho \gtrsim \Gamma_{\rm{osc}} \mu_{\rm{g}} C_{\rm{G}} N .
\end{equation}
As the inequality becomes violated for the case of low Q-factor, the oscillator can no longer reach its maximum amplitude.  In this case, steady-state is achieved with a lower oscillation amplitude such that the 
\begin{equation}
	|\alpha|^2=\frac{\varrho}{\Gamma_{osc} \mu_{\rm{g}} C_{\rm{G}}} .
\end{equation}

\section{Oscillation Threshold and Remarks }
Thus far we have simply assumed that our system oscillates and determined the conditions under which the assumption is self-consistent, utilizing the aid of an equivalent circuit.  Oscillation relies upon coherence that is maintained between the upper and lower of the two transistor modes.  How can such coherence be understood in the context of system thermodynamics? This brings us to a consideration of an oscillation threshold in which parameters are such that it is thermodynamically advantageous for the system to support oscillation.   Let us first consider a readily understandable case in which this is \textit{not} true, that is, when there is no tunneling from the gate to the drain ($\Gamma=0$): The battery provides a reservoir of particles having a specified temperature $T_{\rm{B}}$ and chemical potential $\mu_{\rm{B}}$.  Coupling of the transistor modes to modes of the gate will eventually bring the gate well into thermal equilibrium with the source, meaning the gate temperature and chemical potential are equal to those of the source.  The populations of the upper and lower transistor modes will differ by a Boltzmann factor. In the transistor normal mode basis, thermal equilibrium corresponds to equal populations of the symmetric and anti-symmetric modes (and therefore zero interaction energy, according to Eq. (\ref{Eq:GateInteraction})).  The fact that the two wells are in thermal equilibrium means that a Bose-condensate is present in the gate; one with a  large population of particles so that its chemical potential is sufficient to compensate for the additional bias potential of the source.  With an initial state of population inversion, the formation of the condensate must be accompanied by reverse current flow from gate to source to remove energy from the gate.  

At equilibrium, note that one would expect the dipole mode of the gate condensate to undergo thermal excitation corresponding to $k_{\rm{B}}T_{\rm{s}}$ of energy.  Such dipole oscillation will induce coherence between the transistor modes (because the oscillation frequency is degenerate with the mode spacing) but coupling of the transistor modes with other system modes, including those of the battery, will cause the coherence to decay. In particular, let us suppose that the dipole oscillation of the gate condensate is in some manner excited to a large amplitude:  the amplitude of oscillation will decay to its thermal equilibrium value.  The decay occurs as the excess energy flows back to the source as heat, due to excitation from the lower to upper transistor modes, and subsequent coupling of the upper mode to the source reservoir.  In other words, even if the gate well is perfectly harmonic, coupling of condensate to the source through the transistor modes leads to finite quality factor $Q$ of the dipole oscillation.  Of course, in thermal equilibrium there is no net particle flow between the source and gate.

How do things change if one allows tunneling from the gate to the drain ($\Gamma > 0$) from the upper transistor level?  The majority of particles entering the gate occupy the lower level, while the particles that leave through the drain must do so from the upper state.  In other words, the drain current removes energy from the system as described in the previous two sections. This causes the gate to become colder than the source and have higher chemical potential as well.  The presence of dipole motion of the gate condensate causes particles to be transferred from the lower to upper state, causing more cooling than there would be were dipole motion not resonant with the transistor mode spacing.

\begin{figure}
\includegraphics[width=\columnwidth]{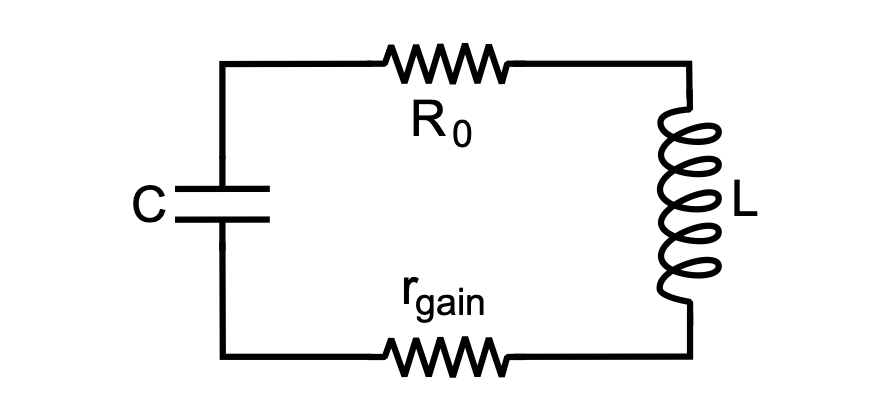}
\caption{\label{fig:oscillator_negative_resistance} The decay of oscillation of an LC circuit due to the presence of a series resistance $R_{\rm{0}}$ can be compensated with the addition of a negative resistor $r_{\rm{gain}}<0$, the latter representing gain provided by other circuit elements (and a source of power).}
\end{figure}

The question, now, is whether the oscillation condition determined earlier decays or is maintained.  In microwave oscillator design it is common to utilize and equivalent oscillator circuit such as the one shown in Fig. \ref{fig:oscillator_negative_resistance}. The combination of the inductor and capacitor support oscillation at a frequency $\omega_{\rm{LC}}=1/\sqrt{L C}$, and an initial excitation will normally decay at a rate embodied by an oscillator resistance $R_{\rm{0}}$.  To compensate for the dissipation and thus maintain oscillation one thinks in terms of adding a negative resistance, $r_{\rm{gain}}<0$. Glauber's inverted oscillator coupled to a reservoir corresponds exactly to such an oscillator with negative resistance in its classical limit \cite{GLAUBER.1986}. In practice this resistance is necessarily nonlinear, being larger in magnitude than the oscillator resistance at small amplitude to provide net gain, and necessarily decreasing to be equal in magnitude to the oscillator resistance at a stead-state amplitude.  We have already associated the tunneling-induced cooling with the negative resistance $r_{\rm{gs}}$ in the equivalent circuit of Fig. \ref{fig:Circuit}, and determined a stead-state value for that resistance (and other parameters). To address the question of oscillation threshold, we are interested in the value of negative resistance relative to an effective resistance that embodies the circuit's dissipative processes.  Nominally we expect cooling, and therefore the magnitude of the negative resistance, to increase as the tunneling to increase.  But at the extreme of large tunneling, population of the upper transistor state in the region of the gate remains small and coherence between the two transistor levels will not be established. 

A full many-body treatment of the transistor to determine dissipation and gain characteristics is significantly cumbersome. A semi-classical kinetic treatment to determine negative resistance characteristics has, however, been carried out in \cite{caliga2012} and is explicitly depicted in Fig. 3 of that work.  That the presence of negative resistance can lead to power amplification using a triple-well atomtronic transistor is established in \cite{caliga2016b}.  The latter work in particular establishes that a transistor circuit can be configured to display the net gain required for oscillation.  

The presence of oscillation, when parameters support it, is thermodynamically advantageous because the system can maximally dissipate the power stored in the battery into the vacuum. This is also the case of the laser, and essentially any driven oscillator. By doing so, it can maximize the rate at which the system approaches thermal equilibrium, in this case with the vacuum of the drain.

It can thus be understood that the oscillation of the condensate is driven by the available power from the battery through the coupling of the two established by the transistor modes.  We began by writing that the combined gate-transistor state as a direct product, Eq. (\ref{Eq:GateStateDirectProduct}). Coherence develops between the upper and lower transistor modes through coupling of the oscillating condensate described by the interaction energy Eq. (\ref{Eq:GateInteraction}). Thus coupling is rather like that between an electromagnetic field and two-level atoms participating in laser action, though in our case the underlying particle modes interact strongly with each other, unlike electromagnetic modes. The transistor mode coherence is manifested in the fact that one of the normal modes is driven to its vacuum state while the other develops a coherent state having finite amplitude.  Self-consistency indicates the gate condensate undergoes dipole oscillation with amplitude given by the number of oscillator levels $N$, $\alpha \approxeq \sqrt{N}$. The dipole oscillation and the oscillation associated with the surviving normal mode are mutually coherent --in particular both oscillate at the gate frequency $\omega_0$ and, in the limit of large gain, are $90^{\circ}$ out of phase. 

\section{Coherent Matterwave Emission}
An electronic oscillator appropriately coupled to the vacuum using an antenna produces a coherent state of the radiation field.  To what extent is there an analogy with the atomtronic oscillator?  We have shown that the oscillator circuit develops a drain current that can be expressed in terms of a gain relative to the uncoupled case.  That gain arises because of the spectrum of particle energies that can undergo resonant tunneling is broadened by the interaction of the incoming particles with the oscillating condensate.  If one considers incoming modes as though they were entering the gate from free space, it would be evident that the current is comprised of normal mode components whose kinetic energy is tied to their phase according to Eq. (\ref{Eq:MatchedPhase}).  The total energy, kinetic plus potential, is of course fixed at $E_{\rm{T}}=(N+2)\hbar\omega_0$.  Nevertheless there remains the fact that the potential energy of the particles varies synchronously with the oscillation of the condensate.  

The previous two sentences are easy enough to say, but sufficiently unfamiliar that it may be worth adding some context:  Consider an ensemble of Bose-condensed particles residing in the ground state of a harmonic oscillator.  We are comfortable with the statement that the system consists of a collection of identical particles, each whose potential and kinetic energies vary \textit{in space} in such a way that its total energy is fixed (at the ground-state energy). In our case, the current is comprised of particles that all have the same total energy, but whose potential and kinetic energy varies \textit{in time}.  This suggests we introduce a wave field associated with the potential, which, just on the vacuum side of the gate-drain barrier we can write as:
\begin{equation}\label{Eq:PotentialWave}
	{\mathcal F}\left(z,t\right)={\mathcal F}_0 \cos\left(k_{\rm{m}} z - \omega_0 t\right), 
\end{equation}
along with another wave field associated with the current:
\begin{equation}\label{Eq:CurrentWave}
	{\mathcal I}\left(z,t\right)={\mathcal I}_0 \cos\left(k_{\rm{m}} z - \omega_0 t\right) ,
\end{equation}
Note in particular that these waves oscillate at the gate frequency rather than with the frequency given by the particle energy, $\omega_{\rm{d}}\equiv E_{\rm{SD}}/\hbar$. This is because the timing, say, of the field maxima, are tied to the oscillation frequency of the gate condensate, and nothing downstream from the gate within the drain can change that timing. 

The two waves of Eqns. (\ref{Eq:PotentialWave}) and (\ref{Eq:CurrentWave})) are analogous to the voltage and current waves associated with a transmission line, and also somewhat analogous to the electric and magnetic fields of an electromagnetic wave (without the vector nature of the latter).  We can make good use of an electromagnetic analog by designating the wave amplitudes to have dimensions of acceleration for ${\mathcal F}$ and momentum for ${\mathcal I}$. The two wave amplitudes are related through a real-valued impedance ${Z}$: 
\begin{equation}
	{\mathcal F}_0 =   { Z}{\mathcal I}_0, 
\end{equation}
where
\begin{equation}
	Z=n^2 \frac{\omega_{0}}{2m}\equiv n^2 Z_0,
\end{equation}
and $n$ plays the role of an index of refraction:
\begin{equation}
	n \equiv \sqrt{\frac{\omega_0}{\omega_d}}.
\end{equation}
We presume the waves propagate at the group velocity of a deBroglie wave:
\begin{equation}
	v_{\rm{m}}=\frac{1}{n}\sqrt{\frac{ 2\hbar \omega_{\rm{0}}}{  m }},
\end{equation}
and are thus associated with the wave number:
\begin{equation}
	k_{\rm{m}}=n\sqrt{\frac{ m \omega_0 }{2 \hbar}}	.
\end{equation}.

The power transmitted into the drain by these waves is:
\begin{equation}
		\left<P_{\rm{d}}\right>=\frac{1}{2}{\mathcal F}_0 {\mathcal I}_0.
\end{equation}
This needs to be consistent with the power supplied by the oscillator to the drain:  
\begin{equation}\label{Eq:Power}
	P_{\rm{d}}=I_{\rm{d}} \hbar \omega_0=V_0 \mu_{\rm{B}} C_{\rm{S}} \omega_0.
\end{equation}
We thus obtain:
\begin{equation}
	{\mathcal I}_0 =\frac{2}{n}\sqrt{m \hbar I_{\rm{d}}}.
\end{equation}
Note that the total power delivered to the drain is, in fact, higher than that specified by Eq. (\ref{Eq:Power}):
\begin{equation}
	P_{\rm{Tot}}=I_{\rm{d}} \hbar \omega_{\rm{d}},
\end{equation}
since a portion of the power is provided not by the battery but from the bias, $V_{\rm{SS}}$.  

\begin{figure}
\includegraphics[width=\columnwidth]{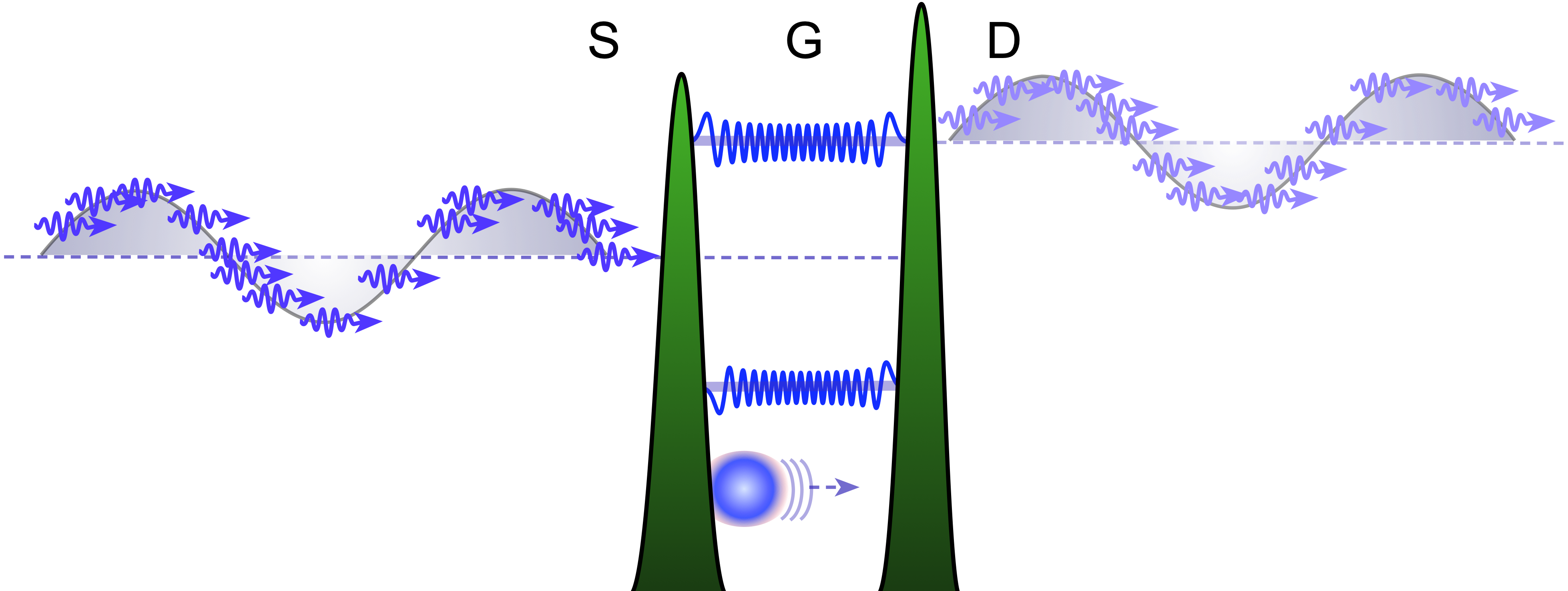}
\caption{\label{fig:CoherentEmission} Coherent matter wave emission from the transistor arises because the resonant tunneling energy of particles arriving at the gate is synchronized to the phase of the oscillating condensate.  The particle flux itself is random.   }
\end{figure}

What, exactly, does the wave of Eq. (\ref{Eq:PotentialWave}) (or Eq. (\ref{Eq:CurrentWave})) represent? It certainly is  \textit{not} a wavefunction describing the quantum state of an individual, or an ensemble, of the associated particles, any more than a classical electromagnetic wave represents a wavefunction. Wikipedia's definition of ``matter wave" (two words) relates directly to the deBroglie wavelength associated with particles having a specific momentum, and more generally in the literature it is used to refer to a wavefunction associated with a massive, usually single, particle.  A deBroglie matterwave will exhibit standing-wave interference, for example, when reflected from a potential having a substantially sharp boundary. The interference will exhibit nodes with a spacing of half of the deBroglie wavelength: $\lambda_{\rm{d}}=n \lambda_{\rm{m}}$ for the typical situation in which $n$ is much less than unity.  On the other hand our wave will also exhibit standing-wave interference if retro-reflected from a barrier, but with a substantially larger distance between nodes: $\lambda_{\rm{m}}/2=\pi/k_m$.  In particular, as impedance is decreased the distance between nodes in deBroglie interference becomes smaller whereas that of the wave of Eq. (\ref{Eq:PotentialWave}) becomes larger.

While it is not a deBroglie wave, it certainly seems appropriate to refer to Eq. (\ref{Eq:PotentialWave}) (or Eq. (\ref{Eq:CurrentWave})) as a ``matterwave" since it is a wave that is associated with massive particles, and in general it will interact with other massive particles. When the distinction is essential, let us refer to these new waves as \textit{classical} or \textit{classically coherent} matterwaves, a phrase that seems particularly appropriate given their connection with the classical equivalent circuit of Fig. \ref{fig:Circuit}. 

The thought of matterwave in classical terms leads one to reconsider the nature of matterwaves, at least in the atomtronics context:  The quantization of the electromagnetic field leads to the powerful notion of a photon, the smallest unit of energy, the ``quantum", carried by a single mode of the electromagnetic field, along with the identification of the photon as a particle. Following the same quantization procedure as for the electromagnetic field one can identify the quantized single-mode excitations of the matterwave field.  As far as we know, such an identity does not exist in the literature, so we will refer to the single-mode excitation quantum of the matterwave field as a ``matteron".  The coherent excitation of the single mode matterwave field is thus our matterwave, and a matterwave is comprised of matterons having energy $\hbar\omega_0$. 

\begin{figure}
\includegraphics[width=\columnwidth]{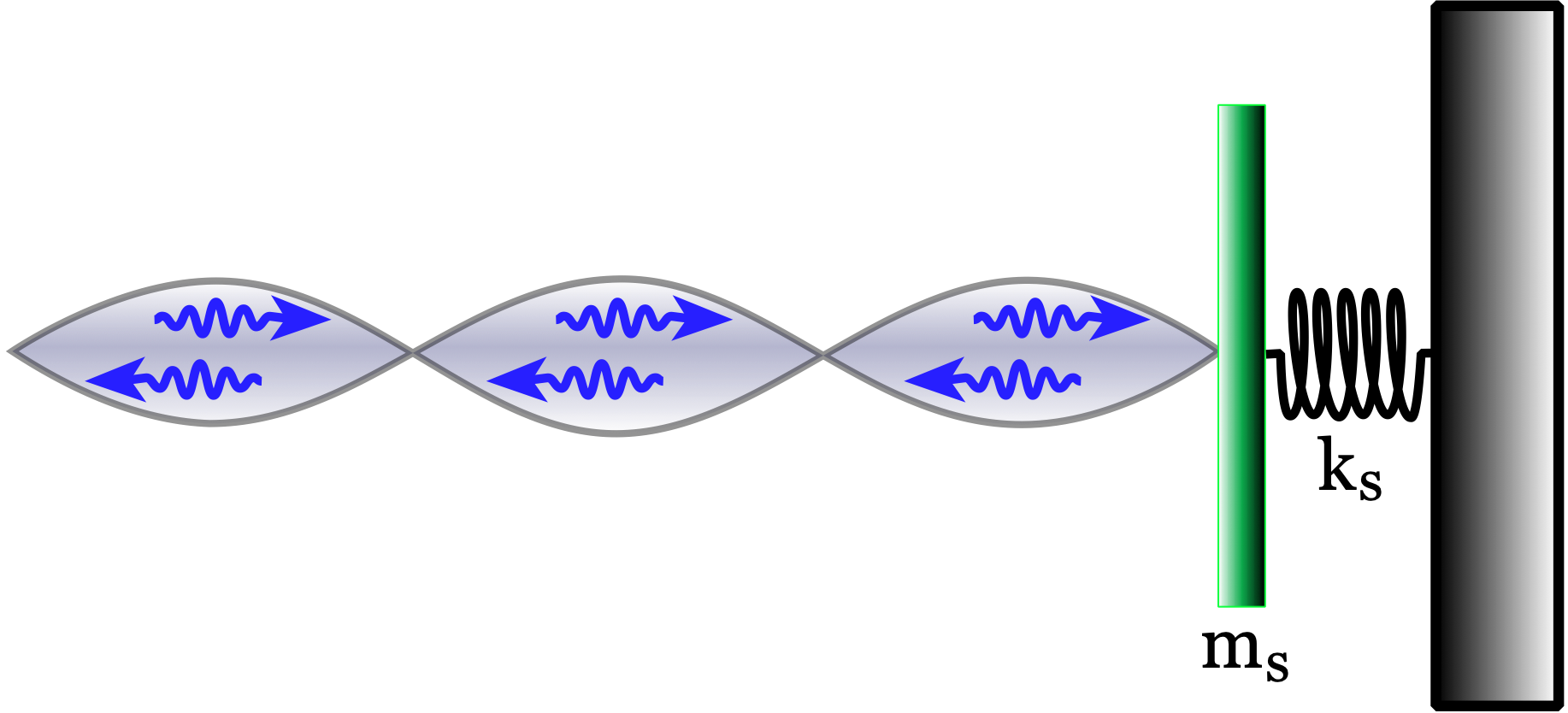}
\caption{\label{fig:MatteronDetector} A mass-spring combination serves as a matterwave detector for a wave frequency given by the resonant frequency of the mechanical oscillator, $\omega_{\rm{s}}=\sqrt{k_{\rm{s}}/m_{\rm{s}}}$. The standing wave node spacing is $\lambda_{\rm{m}}/2$ and in particular is different from that of the associated atoms' deBroglie wavelength.}
\end{figure}

The term may at first seem superfluous because, after all, we know the drain output from our atomtronic circuit is comprised of atomic particles.  But the quanta associated with the emitted matterwave field are not atoms. Consider the following Gedanken experiment to detect matterons and not atoms \textit{per se}:  Fig. \ref{fig:MatteronDetector} illustrates a matterwave mirror of mass $m_{\rm{m}}$ attached to a spring having spring constant $k_{\rm{s}}$, in turn attached to a wall.  The mass-spring system will have resonant frequency $\omega_{\rm{s}}=\sqrt{k_{\rm{s}}/m_{\rm{s}}}$ and let us suppose the losses are low.  We expose the mechanical oscillator to a single frequency matteron beam for a significant time compared with its oscillation period, then subsequently measure its oscillation amplitude.  Energy will have been deposited in the detector provided $\omega_0\approx\omega_{\rm{s}}$. The momentum associated with the matteron is evidently:
\begin{equation}
	p=\hbar k_0 = \sqrt{2 m \hbar\omega_0 }.
\end{equation}
Importantly, atom number is conserved in interactions, but matteron number is generally not, and while matterons carry momentum, their mass is zero.  While the analogy is instructive, whether the concept of matterons proves to be as useful as that of photons remains to be seen.  

\section{Conclusions}
Our analysis of the transistor oscillator began with the ansatz that the gate many-body wavefunction is a displaced ground state, which we have referred to as a ``massive coherent state". The idea is to justify the ansatz based on self-consistency of the steady-state solution. The interesting circuit dynamics is largely governed by the interaction energy, Eq. (\ref{Eq:GateInteraction}), between the gate and the transistor modes. Recognizing that the gate well has a finite number of energy levels, we find that the oscillation amplitude of the gate condensate must saturate, i.e., the gate many-body wavefunction is not a perfect displaced ground state, but nevertheless it is a condensed state and furthermore there is a maximum oscillation amplitude that corresponds to a maximum interaction energy.  The maximum interaction energy, in turn, leads to a drain current that is substantially larger than it would be in absence of the gate coupling.  The drain current induces cooling of the gate particles.  That cooling, we argue, maintains the gate particles in their condensed state, at least if the anharmonicity of the gate well is not overly large.  The final link in self-consistency is apparent in the interaction energy, whose average value is driven to zero only as the average relative phase between the oscillating condensate and the (single) normal mode reaches $\pi/2$. This occurs at a maximum of the oscillation amplitude, and is also characteristic of any forced simple oscillator.

We transition from a microscopic quantum picture to a classical one utilizing an equivalent circuit that is  similar to that which would be used to analyze a transistor oscillator circuit such as the Colpitts oscillator of Fig. \ref{fig:ColpittsFET}. The steady-state circuit dynamics is fully characterized in terms of the circuit design parameters in conjunction with the battery potential and temperature.  Of particular note is the determination of the atomtronic analogs to the electronic transistor transconductance and current gain.  It is fair to suppose, however, that the predictions of the key parameters are valid provided the battery chemical potential is not too large compared with $\hbar \omega_0$.
Our treatment of the oscillator in terms of an equivalent circuit led us to contemplate classical coherent matterwaves as analogs of electromagnetic waves.  The coherence of the matterwaves is a consequence of the phase-dependent potential arising from the transistor-gate interaction energy.  The phase dependence translates to a flow of particles whose kinetic and potential energies oscillate in time in such a way that their total energy remains fixed.  Perhaps counter-intuitively, the particle flux does not oscillate in time, yet this is simply the same as the photon flux associated with a coherent electromagnetic wave. We have seen that the classically coherent matterwave is quite distinct from the elemental deBroglie matterwaves associated with massive particles. As one moves from a region of high atomic potential energy to a region of low potential for example, the deBroglie wavelength becomes shorter, whereas the classical matterwave wavelength becomes longer.  In the circuit context, classical matterwaves can be treated much like the electromagnetic waves of microwave circuitry.  That is, their behavior is determined by their frequency and the impedance properties of the circuitry.  

The theory presented here has sought to extract qualitative characteristics from what is otherwise a substantially complicated problem in many-body quantum thermodynamics.  Our treatment gives rise several predictions; most notable is the emission of a classical matterwave, which can be verified using interferometry. Another signature of oscillation is the presence of a small condensate in the gate accompanied by heating of the source well, (due to power dissipation by the battery's internal resistance), in contrast to a large condensate reflecting thermal equilibrium and net cooling of the source well resulting from the most energetic particles escaping to the vacuum.  Other predictions can be experimentally tested by varying circuit parameters, for example, to establish the threshold and tuning characteristics of the circuit.

Our work has emphasized the analogs and duals between atomtronics and electronics in an attempt to move atomtronics to a similar footing as electronics.  Electronics as a discipline has managed to codify incredibly complicated non-thermal equilibrium dynamical systems through a relatively small set of principles, rules, and heuristics.  In this way we hope to leverage the vast knowledge within electronics to bring useful quantum advantage to a technologically relevant foreground.  

\begin{acknowledgments}
The author would like to thank S. Du, K. Krzyzanowska, J. Combs and D. Gu{\'e}ry-Odelin for fruitful discussions.
\end{acknowledgments}

\bigskip
 
\bibliography{Transistor_Oscillator}

\end{document}